\documentclass[letterpaper, 10 pt, conference]{ieeeconf}  
\IEEEoverridecommandlockouts           

\usepackage{cite}
\usepackage{amsmath,amssymb,amsfonts}
\usepackage{graphicx}
\usepackage{textcomp}
\usepackage{xcolor}
\def\BibTeX{{\rm B\kern-.05em{\sc i\kern-.025em b}\kern-.08em
    T\kern-.1667em\lower.7ex\hbox{E}\kern-.125emX}}
\usepackage[utf8]{inputenc}
\usepackage{mathrsfs}
\usepackage{xcolor}
\usepackage{mathtools}
\usepackage{enumerate}
\usepackage{booktabs}
\usepackage{romannum}
\usepackage{float}
\usepackage{algorithm}
\usepackage{dirtytalk}
\usepackage[normalem]{ulem} % for \sout
\usepackage{amsthm}
\newtheorem{theorem}{Theorem}

\usepackage[normalem]{ulem}

\newcommand{\x}[1]{\mathbf{x}_{#1}}

\newcommand{\e}[1]{\mathbf{e}_{#1}}

\makeatletter
\renewcommand*\env@matrix[1][\arraystretch]{%
  \edef\arraystretch{#1}%
  \hskip -\arraycolsep
  \let\@ifnextchar\new@ifnextchar
  \array{*\c@MaxMatrixCols c}}
\makeatother

\newcommand{\argminD}{\arg\!\min} % AlfC
\newcommand{\argmaxD}{\arg\!\max} % AlfC

\newcommand{\finite}{\star}

\newcommand{\RR}{{\mathbb{R}}}

\newcommand{\sgn}{\mathop{\mathrm{sgn}}}
\DeclareUnicodeCharacter{2212}{-}
\title{\Large\bf Guarding a Non-Maneuverable Translating Line with an Attached Defender}
\author{Goutam Das$^{1}$, Michael Dorothy$^{2}$, Zachary I. Bell$^{3}$, and Daigo Shishika$^{4}$% <-this % stops a space
\thanks{We gratefully acknowledge the support of ARL grant ARL DCIST CRA W911NF-17-2-0181. The views expressed in this paper are those of the authors and do not reflect the official policy or position of the United States Government, Department of Defense, or its components.}% <-this % stops a space
\thanks{$^{1}$Goutam Das, PhD student, Electrical and Electronics Engineering, George Mason University, 4400 University Dr, Fairfax, VA 22030, USA
        {\tt\small gdas@gmu.edu}}%
\thanks{$^{2}$Michael Dorothy. Army Research Directorate, DEVCOM Army Research Laboratory, APG, MD.
        {\tt\small michael.r.dorothy.civ@army.mil}}%
\thanks{$^{3}$Zachary I. Bell is with the Munitions Directorate, Air Force Research Laboratory, Eglin AFB, FL 32542, USA {\tt\small zachary.bell.10@us.af.mil.}}%
\thanks{$^{4}$Daigo Shishika, Assistant Professor, Department of Mechanical Engineering, George Mason University,
        4400 University Dr, Fairfax, VA 22030, USA
        {\tt\small dshishik@gmu.edu}}%
}
\begin{document}
\maketitle 
\thispagestyle{empty}
\pagestyle{empty}

%%==============================================================================
\begin{abstract}
% In this paper, we consider a target-guarding differential game where the defender must protect a linearly translating line segment by intercepting the attacker who tries to reach it. In contrast to common target-guarding problems, we assume that the defender is attached to the target and moves along with it. This assumption affects the defender's maximum speed 
% % \mike{in absolute coordinates, which depends on the target's} 
% in absolute coordinates, which depends on the target's heading direction.
% Zero-sum differential game of degree for both the attacker-winning and defender-winning scenarios are studied, where the payoff is defined to be the distance between the two agents at the time of game termination.
% We derive the equilibrium strategies and the Value function by leveraging the solution for the infinite-length target scenario. 
% The zero level set of this Value function provides the barrier surface that divides the state space into defender winning and attacker winning regions.
% We present simulation results to demonstrate the theoretical results.
 In this paper we consider a target-guarding differential game where the defender must protect a linearly translating line-segment by intercepting an attacker who tries to reach it. 
 In contrast to common target-guarding problems, we assume that the defender is attached to the target and moves along with it. This assumption affects the defender’s maximum speed in inertial frame,
%  in absolute coordinates, 
 which depends on the target’s 
%  heading direction. 
 direction of motion.
 Zero-sum differential {game of degree} for both the attacker-win and defender-win scenarios are studied, where the payoff is defined to be the distance between the two agents at the time of game termination. We derive the equilibrium strategies and the Value function by leveraging the solution for the infinite-length target scenario. 
 The zero-level set of this Value function provides the barrier surface that divides the state space into defender-win and attacker-win regions. We present simulation results to demonstrate the theoretical results.
\end{abstract}

\section{INTRODUCTION}

Pursuit-evasion games (PEG) are a class of differential games in which an agent (i.e., pursuer/defender) attempts to capture another agent (i.e., evader/attacker) who seeks to avoid or delay the capture. 
% Various applications of this type of problem have been studied by the robotics and controls community, including missile guidance \cite{Garcia2017}, aircraft defense \cite{Fang2018, Vitaly2010}, robot navigation\cite{Guilamo2004}, and autonomous vehicles \cite{Wang2021}, just to name a few.
This paper focuses on a particular variant of PEG that involves 
an asset/target
% a non-maneuverable translating line 
that must be guarded.
% Such a scenario has high relevance to both civilian and military defense applications.
In both civilian and military defense applications\cite{Garcia2017,Fang2018, Vitaly2010,Guilamo2004, Wang2021}, such a scenario is highly relevant.

Target-attacker-defender games (TADG) study situations where the attacker seeks to reach the target without being intercepted by the defender.
In the literature, targets are typically modeled as points or agents that are stationary \cite{Selvakumar2021} and guarded by the defenders. Alternatively, the target can be a non-stationary agent which cooperates with the defender by either actively evading the attacker or by rendezvousing with the defender \cite{Liang2019,Liang2020,Liang2021}.
The defender wins the game either by intercepting the attacker \cite{Garcia2017, Garcia2018, Garcia2019, Rubinsky2014} or by rendezvousing with the target \cite{OYLER2016}.

% A related class of PEG is called target guarding (TG), which was first introduced by Isaacs \cite{Issacs1965}.
% The main difference between TG and TAD games is the target is a region in TG games instead of a point, which makes the rendezvous-type strategy invalid for the defender.
% There are many different variants of TG, including \emph{reach-avoid} games \cite{Zhou2012,Huang2011,Chen2014} and \emph{coastline guarding} or \emph{border-defense} problems \cite{Garcia_Coastline2019,VonMoll2020BD,Garca2020TheBS}.
% These works extend the problem to multi-agent scenarios and consider different geometric settings; however, it is generally assumed that the agents have simple motio  and can freely move on a planar region.

A related class of PEGs is target guarding (TG), which was introduced by Isaacs \cite{Issacs1965}.
In TG games, the target is a region rather than a point, which renders rendezvous-type strategies ineffective for the defender.
% The main difference between TG and TAD games is the target is a region in TG games instead of a point, which makes the rendezvous-type strategy invalid for the defender.
% There are many different variants of TG, 
Several variants of TG exist, including \emph{reach-avoid} games \cite{Zhou2012,Huang2011,Chen2014} and \emph{coastline guarding} or \emph{border-defense} problems \cite{Garcia_Coastline2019,VonMoll2020BD,Garca2020TheBS}.
% These works extend the problem to multi-agent scenarios and consider different geometric settings; however, it is generally assumed that the agents have simple motion and can freely move on a planar region.
These works have extended the problem to multi-agent scenarios and considered various geometric settings; however, it is generally assumed that agents have simple motions and are free to move within a planar space.

In this study, we are interested in TG scenarios where the defender is constrained to move only along the perimeter of the target.
Similar works have been previously studied as \emph{perimeter-defense} games \cite{VonMol2020, Shishika2020review,Shishika2020}.
Unlike standard TG, these papers assume that the defender cannot pass through the target region. 
Therefore, the defender must move around the perimeter in order to reach the attacker, thereby affecting the dynamics and thus the capturability.
Different variants have been studied with differential game techniques \cite{VonMol2020} and with geometric approaches \cite{Shishika2020review,Shishika2020}; however, these studies are based on stationary target regions.

% In this paper, 
% we consider a defender that is constrained to move on a linear target that translates on a plane with a constant speed. 
In this paper, we consider a target that translates on a plane.
{As an initial step towards a more realistic scenario, a non-maneuverable target with no rotational motion is studied.} 
% non-maneuverable linear target that translates on a plane with a constant speed. 
% Note that, the target motion is restricted to only translation in the space and the target has no rotation in inertial frame. 
% This paper generalizes the strategies of the players for any given translation direction of the target instead of translating only towards the positive $x$-axis \cite{Goutam2022}.
% It is important to note that the target motion is limited to translation in space and does not involve rotation in an inertial frame.
The attacker moves freely and tries to reach the target while avoiding the defender. However, the defender is constrained to move only on the linear target. 
In the inertial frame, the defender is dragged in the direction of the target's motion, but the attacker is not affected by the motion of the target. 
In this context, there is a connection to the work presented in \cite{Sun2017,Sun2015}, where PEG is played in a flow field;
however, the results in \cite{Sun2017,Sun2015} do not extend naturally to TG objectives considered in this paper. Moreover, the flow field affects only one of the two agents in this paper.

The main contributions of this paper are: 
% (i) generalizing the target's direction of motion,
(i) the characterization of the barrier surface that separates the state space into defender-win and attacker-win regions; and
(ii) the equilibrium strategies and the Value function in each regions.
% for both the attacker-win and defender-win scenarios.
% By considering the target to be a line segment that translates in a given direction, this paper completes the previous work of translating line guarding problem and provides solution for \textit{Game of Degree} for both the attacker-winning and defender-winning game.
By allowing the target to translate in an arbitrary direction, this paper generalizes the result in \cite{Goutam2022} which assumes that the target can only translate in $x$-direction.
In addition, we provide the solution to both the attacker-win and defender-win scenarios, where the latter was missing in \cite{Goutam2022}.

%%===========================================================================
\section{PROBLEM FORMULATION}\label{sec:prob_formulation}
% \vspace{2mm}
\begin{figure}[t!]
    \centering
    % \centerfloat
    \vspace{2mm}
    \includegraphics[width = \columnwidth]{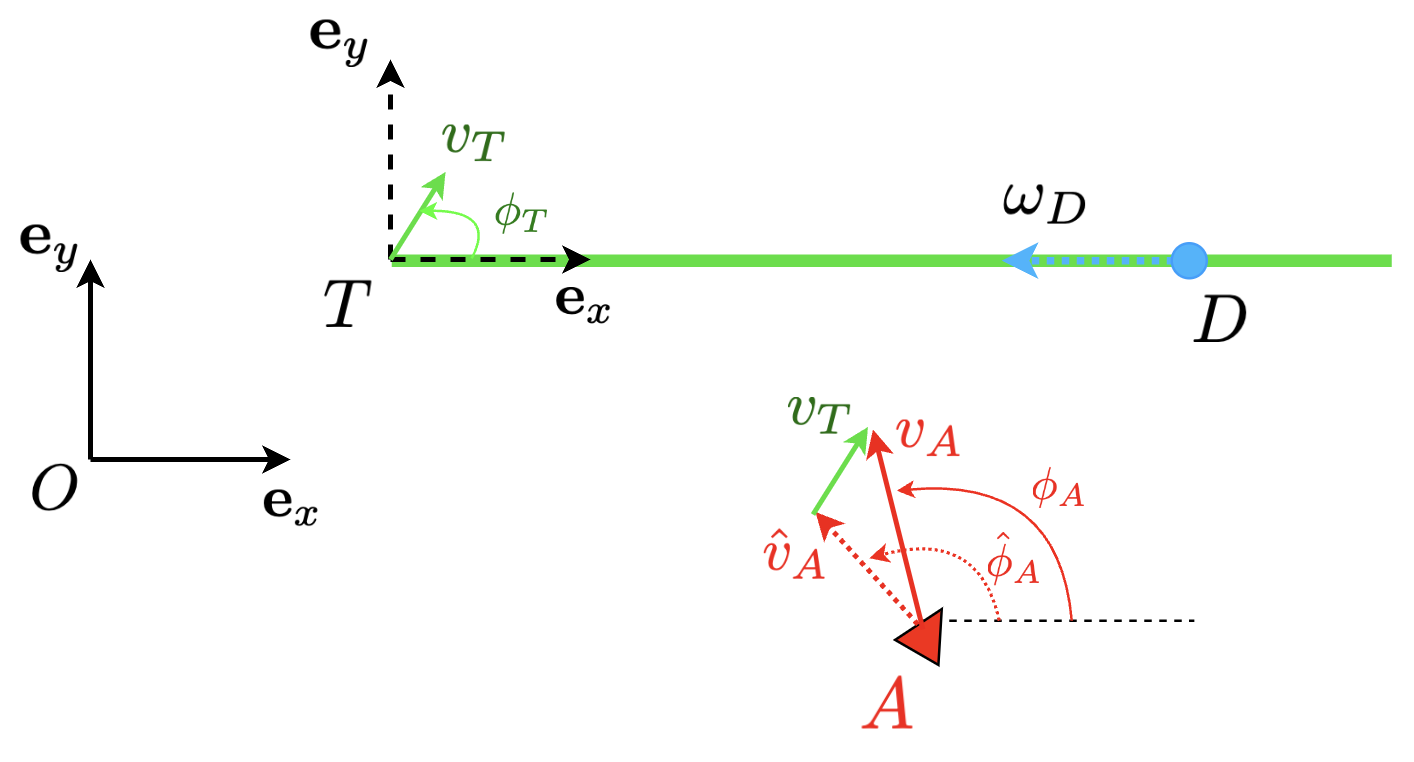}    
    \caption{Illustration of translating line guarding problem for one defender and one attacker in target frame and in inertial frame.
    }
    \label{fig:fig1}
\end{figure}
This section formulates the translating line guarding game on a plane for one defender and one attacker shown in Fig.~\ref{fig:fig1}.
The inertial frame $\mathcal{I} = (O, \mathbf{e}_x, \mathbf{e}_y)$ is defined by the origin $O$, and the basis vectors $\mathbf{e}_x$ and $\mathbf{e}_y$. The positions of the agents in $\mathcal{I}$ are denoted as $\x{i} = \left[ x_i, y_i \right]^\top \in \RR^2$, where $i = \{A,D,T\}$ represents the attacker, defender, and target, respectively. 
% The target is a compact set, $T \subset \RR^2$, is a line segment with length $L$ is aligned with the $x$-axis, with end points at $\x{T}$ and $\x{T} + L \mathbf{e}_x$,
The target, $T$, is a line segment and it is aligned with $\e{x}$. The length of $T$ is $L$, and so the endpoints are given by $\x{T}$ and $\x{T} + L \mathbf{e}_x$.

The dynamics of the attacker in $\mathcal{I}$ are 
\begin{equation}
    \begin{aligned}
        \dot{\x{}}_A = 
        \begin{bmatrix}
        \dot{x}_A \\
        \dot{y}_A
        \end{bmatrix}
        =         
        v_A \begin{bmatrix}
        \cos{\phi_A} \\
        \sin{\phi_A}
        \end{bmatrix},
    \end{aligned}
    \label{eq1}
\end{equation}
where $\phi_A \in [-\pi,\pi]$ is the attacker's control (i.e., its heading angle), and $v_A$ is its speed given as part of the game parameters.
The target moves at a constant velocity
\begin{equation}
    \begin{aligned}
        \dot{\x{}}_T = 
        \begin{bmatrix}
        \dot{x}_T \\
        \dot{y}_T
        \end{bmatrix}
        =         
        v_T \begin{bmatrix}
        \cos{\phi_T} \\
        \sin{\phi_T}
        \end{bmatrix},
    \end{aligned}
    \label{eq2}
\end{equation}
where $v_T$ and $\phi_T$ are the game parameters known to the players. 
% the constant speed and direction of the target velocity, respectively.
The defender is assumed to be \say{attached} to the target, and can move in the $x$-direction relative to the target:
\begin{equation}
    \begin{aligned}
        \dot{\x{}}_D = 
        \begin{bmatrix}
        \dot{x}_D \\
        \dot{y}_D
        \end{bmatrix}
        =         
        v_T \begin{bmatrix}
        \cos{\phi_T} \\
        \sin{\phi_T}
        \end{bmatrix} 
        + \begin{bmatrix}
        \omega_D \\
        0
        \end{bmatrix},
    \end{aligned}
    \label{eq3}
\end{equation}
where $\omega_D \in [-1, 1]$, is the defender's control.
% Note that since the defender is attached to the target, $\x{D} \in [\x{T}, \x{T}+L\mathbf{e}_x]$ and is therefore restricted to moving on the target.
Since the defender is attached to the target, its states must satisfy $\x{D} \in [\x{T}, \x{T}+L\mathbf{e}_x]$.
Consequently, 
% $w_D \geq 0$ when the defender is at the leftmost point on the target and $w_D \leq 0$ when the defender is at the rightmost point of the target.  
$\omega_D \geq 0$ when $\x{D}=\x{T}$, and $\omega_D \leq 0$ when $\x{D}=\x{T}+L\mathbf{e}_x$.  

For convenience, we perform our analysis in the translating target frame $\mathcal{B} = (T, \mathbf{\e{}}_x, \mathbf{\e{}}_y)$ attached to the leftmost point of the target. Let $\hat{\x{}}_i, i = \{A, D\}$ denote the agents' positions in $\mathcal{B}$, where $\hat{y}_D = 0$. 
Letting $\hat{\x{}} = \left[\hat{x}_D, \hat{x}_A, \hat{y}_A \right]^\top$ 
represent the stacked state and using \eqref{eq1}--\eqref{eq3} yields 
\begin{equation} \label{eq: kinematics}
    \begin{aligned}
        f({\hat{\x{}},\omega_D, \phi_A}) = 
        \dot{\hat{\x{}}} = 
        \begin{bmatrix}
        \dot{\hat{x}}_D \\
        \dot{\hat{x}}_A \\
        \dot{\hat{y}}_A
        \end{bmatrix}
        =         
        \begin{bmatrix}
        \omega_D \\
        v_{A}\cos{\phi}_A-v_{T}\cos{\phi}_T \\ 
        v_{A}\sin{\phi}_A-v_{T}\sin{\phi}_T
        \end{bmatrix}.
    \end{aligned}
\end{equation}

Assume the following on the agents' speeds: 
\begin{itemize}\setlength{\itemindent}{1em}
    \item[\textbf{A1})] The attacker is faster than the target, $v_A>v_T$.
    % \item[\textbf{A2})] The defender is faster than the attacker, $-1+v_T>v_A$.
    
    \item[\textbf{A2})] The defender can outrun the attacker in $\e{x}$ direction: i.e.,
    $v_A < 1-|v_T\cos{\phi}_T|$.
    % 
    % \item[\textbf{A3})] Without loss of generality, the attacker starts below the target, $\hat{y}_A<0$. 
\end{itemize}
% \sout{Assumption (A1) ensures that the attacker can reach the target in case of pure pursuit.}
Assumption (A1) avoids the degenerate case where the attacker is too slow to reach the target even if there was no defender.
% Assumption (A2) ensures the attacker's maximum speed is bounded by the defender's speed, $v_D \in [-1+v_T,1+v_T]$. Note the defender wins the game if $\hat{x}_D = \hat{x}_A$ since the defender is faster than the attacker.
%
% Assumption (A2) limits the maximum attacker speed in $\mathcal{I}$ and as a result once the defender is aligned with the attacker defender has sufficient control authority to maintain that irrespective of the attacker control.
% Using assumption (A2), the attacker's maximum speed is alwasy less than the defenders maximum speed in $/mathcal{I}$, and as a consequence, once the defender has aligned itself with the attacker, the defender has sufficient control authority to maintain that regardless of the attacker's control.
%
% \daigo{
Assumption (A2) ensures that once the defender aligns itself with the attacker (i.e., $\hat{x}_D = \hat{x}_A$), it has sufficient control authority to maintain that alignment regardless of the attacker's control (as long as $\hat{x}_A\in[0,L]$).
% }
% Based on assumption (A2), the attacker's speed is always less than the defender's speed in $\mathcal{I}$. As a consequence, once the defender aligns itself with the attacker (i.e., $\hat{x}_D = \hat{x}_A$), it has sufficient control authority to maintain that alignment regardless of the attacker's control.

% Therefore, we consider $\hat{x}_A = \hat{x}_D$ a part of the terminal condition (i.e., the defender has successfully intercepted the attacker and thwarted the attack). 
%
For this paper, we consider the \emph{game of kind} as the question of whether the attacker can reach the target 
% with a non-zero \textit{miss-distance} from the defender (i.e., attacker wins the game), 
or if the defender can prevent it.
The barrier surface that provides the answer to this question will be obtained by solving a related \emph{game of degree}.
% \daigo{
The terminal conditions and the payoff functions that define the {game of degree} will be provided separately for the attacker-win and defender-win scenarios.
% }
% capture the attacker by matching its $x$-coordinate (i.e., $x_A = x_D$, which implies that the defender wins since the attacker is slower than the defender).
% \sout{In the following sections, the terminal conditions  for the attacker-win and defender-win scenarios are provided. Additionally, the barrier surface that separates these two cases is derived by solving a related \textit{Game of Degree}.}

%%========================================================================
\section{ATTACKER-WIN SCENARIO}\label{sec: att-win}

In this section, we are concerned with the {game of degree} when the attacker is able to reach the target (i.e., drive $\hat{y}_A \xrightarrow{} 0$), with a nonzero \emph{miss-distance} from the defender. The initial condition of the system lies inside the attacker-win region (i.e., $\hat{\x{}} \in \mathcal{R}_A)$, and we use subscript $a$ to refer the {game of degree} in this region. 

We consider a zero-sum differential game with the following payoff that describes the {miss-distance} \begin{equation}\label{payoff}
    \begin{aligned}
    J_a(\hat{\x{}}_0,\omega_D,\phi_A) \hspace{-2pt}&=\hspace{-2pt} \Phi_a(\hat{\x{}}_f) \hspace{-2pt}=\hspace{-2pt}|{\hat{x}_A(t_f) - \hat{x}_D(t_f)}|,
\end{aligned}
\end{equation}
where $\hat{\x{}}_f:=\hat{\x{}}(t_f)$ and $t_f$ represent the terminal time.
Here the defender is the minimizing player who seeks to minimize the {miss-distance}, and the attacker is the maximizing player whose goal is to maximize it. 
If an equilibrium exists, the value function is defined as
\begin{align}
    V_a(\hat{\x{}}_0) &= \min\limits_{\omega_D} \max\limits_{\phi_A} J_a = \max\limits_{\phi_A} \min\limits_{\omega_D} J_a.
\end{align}
The equilibrium strategies $\omega_D^*$, $\phi_A^*$ satisfy the following saddle-point condition:
\begin{align}
    J_a(\cdot,\omega_D^*,\phi_A) \leq J_a(\cdot,\omega_D^*,\phi_A^*) \leq J_a(\cdot,\omega_D,\phi_A^*).
\end{align}

The terminal constraint is given by
\begin{align}\label{eq: terminal_cond_RA}
  \psi_a(\hat{\mathbf{\x{}}}_f) = \hat{y}_A(t_f) &= 0.
\end{align}
Thus, the terminal surface is defined by the set of states satisfying \eqref{eq: terminal_cond_RA}:
\begin{equation}
    \begin{aligned}
    \mathcal{S}_{T_a} = \{\hat{\x{}} \mid \hat{y}_A = 0 \text{ and, } \hat{x}_A \in [0,L]\}.
    \end{aligned}
\end{equation}
We will derive $V_a$ and the corresponding equilibrium strategies in the following sections.
%%======================================
% \subsection{Infinite Length Target}\label{subsection: Infinite Length Target}
%%======================================
% As a building block towards the complete solution, in addition to (A1) - (A3) we make the following assumption:
% \begin{itemize}\setlength{\itemindent}{1em}
%     \item[\textbf{A4})] The length of the target is infinite.
% \end{itemize}
\subsection{Infinite Length Target}\label{subsection: Infinite Length Target}
As a building block towards the complete solution, 
% \sout{in addition to (A1) - (A3) we also assume} 
this section assumes that the target length is infinite. 
% The system dynamics and terminal constraint remains the same as previous analysis.
The system dynamics, payoff, and terminal constraint remains the same as stated in \eqref{eq: kinematics}, \eqref{payoff}, and  \eqref{eq: terminal_cond_RA} respectively.
% Terminal time, $t_f$, is given by the time the attacker reaches the target, and 
The terminal surface for the infinite target is given by
\begin{equation}
    \begin{aligned}
      \mathcal{S}_{T_a,\text{inf}} = \{ \hat{\x{}} \mid \hat{y}_A = 0\}.
    \end{aligned}
\end{equation}

%%=================================================================================
\subsubsection{First Order Necessary Conditions for Optimality} \label{subsection: Infinite Length Target, FONC}
% In this section we derive
This section presents the optimal strategies for the defender and the attacker for $\mathbf{\hat{\x{}}} \in \mathcal{R}_A$. 
First order necessary conditions \cite{Kirk1970} are used to derive the equilibrium strategies for the players. The solution approach involves defining and optimizing a function known as Hamiltonian. The Hamiltonian for the differential game \eqref{eq: kinematics} is given by
% formulated as
\begin{equation}\label{eq: hamiltonian_RA}
\begin{aligned}
    \mathcal{H}_a&(\mathbf{\hat{\x{}}},\omega_D, \phi_A, \sigma, t) \\
    &=  
     l({\mathbf{\hat{\x{}}},\omega_D, \phi_A},t) + \boldsymbol{\sigma}^\top(t) f(\mathbf{\hat{\x{}}},\omega_D, \phi_A,t) \\
    &= \sigma_{\hat{x}_D} \omega_D + \sigma_{\hat{x}_A} v_A\cos{\phi_A} - \sigma_{\hat{x}_A} v_T\cos{\phi_T} 
    \\
    &\quad +\sigma_{\hat{y}_A} v_A\sin{\phi_A} - \sigma_{\hat{y}_A} v_T\sin{\phi_T},
\end{aligned}
\end{equation}
% 
% where 
% $\mathbf{z}:=\{\hat{\x{}},w_D, \phi_A\}$,  
% \edit{
where the integral cost $l(\cdot)$ is 0 in our problem,
% }
% $l(\cdot)= 0,$ is the integral cost component,
% \sout{integral cost, $L = 0$}, 
and $\boldsymbol{\sigma} := [\sigma_{\hat{x}_D}, \sigma_{\hat{x}_A}, \sigma_{\hat{y}_A}]^\top$, is the adjoint vector.
Notice that the Hamiltonian in \eqref{eq: hamiltonian_RA} is a separable function
of the controls $\omega_D$ and $\phi_A$, and thus \textit{Isaacs’ condition} \cite{Issacs1965} , \cite{Basar2011} holds:
\begin{align}
    \min\limits_{\omega_D} \max\limits_{\phi_A} \mathcal{H}_a = \max\limits_{\phi_A} \min\limits_{\omega_D} \mathcal{H}_a.
\end{align}
The equilibrium adjoint dynamics are given by
\begin{align}\label{eq: adjoint_dynamics}
        \dot{\boldsymbol{\sigma}} &= \frac{\partial \mathcal{H}_a}{\partial \mathbf{\hat{\x{}}}}
        = [0,\; 0,\; 0].
        % \begin{bmatrix}
        % 0 & 0 & 0
        % \end{bmatrix}.
\end{align}
The terminal adjoint values are obtained from the transversality condition \cite{Bryson1975}:
\begin{equation} \label{eq: transversality_cond}
    \begin{aligned} 
    \boldsymbol{\sigma}^\top(t_f) 
    &=\frac{\partial \Phi_a}{\partial \hat{\mathbf{\x{}}}_f} + 
    \eta \frac{\partial \psi_a}{\partial \hat{\mathbf{\x{}}}_f}   
    =  
    % \begin{bmatrix}
    % -\lambda & \lambda & \eta
    % \end{bmatrix},
    [-\lambda,\; \lambda,\; \eta],
\end{aligned}
\end{equation}
 where $\lambda := \sgn(\hat{x}_A - \hat{x}_D)$ and $\eta$ is Lagrange multiplier vector \cite{Kirk1970}.
%  The length of $\eta$ is the same as the number of constraints.
Therefore, with \eqref{eq: adjoint_dynamics} and \eqref{eq: transversality_cond}, the following holds:
% \label{eq: transversality_cond}
\begin{equation} 
    \begin{aligned}
    \boldsymbol{\sigma}(t) &= 
    % \begin{bmatrix}
    % -\lambda & \lambda & \eta    
    % \end{bmatrix},
    [-\lambda,\; \lambda,\; \eta]^\top,
    &\forall \hspace{0.0mm} t \in [t_0, t_f].
\end{aligned}
\end{equation}
The terminal Hamiltonian satisfies 
\begin{equation}
     \begin{aligned}
     \mathcal{H}_a(t_f) = -\frac{\partial \Phi_a}{\partial t_f} -\eta \frac{\partial \phi_a}{\partial t_f} &= 0,
 \end{aligned}
\end{equation}
and $\frac{d\mathcal{H}_a}{dt} = 0$, therefore, $\mathcal{H}_a(t) = 0$ for all $t \in [t_0, t_f]$.

The equilibrium control actions of the attacker and the defender maximize and minimize \eqref{eq: hamiltonian_RA} respectively: $\mathcal{H}_a^* = \text{max}_{\phi_A} \text{min}_{\omega_D} \mathcal{H}_a$. 
For the saddle point solution of the problem, we have
\begin{equation}\label{wd*}
    \begin{aligned}
    \omega_D^* &= 
    \argminD_{\omega_D} \mathcal{H}_a
    \\ 
    &= \argminD_{\omega_D} (\sigma_{\hat{x}_D}\omega_D) 
    = 
    -\sgn({\sigma_{\hat{x}_D}}) = \lambda,
    \end{aligned}
\end{equation}
\begin{equation}\label{phi*}
    \begin{aligned}
    \phi_A^* &= 
    \argmaxD_{\phi_A} \mathcal{H}_a
    \\
    &= \argmaxD_{\phi_A} (\sigma_{\hat{x}_A} v_A\cos{\phi_A} + \sigma_{\hat{y}_A} v_A\sin{\phi_A}).
    \end{aligned}
\end{equation}
Solving (\ref{phi*}), we have
\begin{equation}\label{eq: cosphi*_RA}
    \begin{aligned}
    \cos{\phi_A^*} &= \frac{\sigma_{\hat{x}_A}}{\sqrt{\sigma_{\hat{x}_A}^2 + \sigma_{\hat{y}_A}^2}}  = \frac{\lambda}{\sqrt{\eta^2 + 1}},
    \end{aligned}
\end{equation}
\begin{equation}\label{eq: sinphi*_RA}
    \begin{aligned}
    \sin{\phi_A^*} &= \frac{\sigma_{\hat{y}_A}}{\sqrt{\sigma_{\hat{x}_A}^2 + \sigma_{\hat{y}_A}^2}} = \frac{\eta}{\sqrt{\eta^2 + 1}}
    \end{aligned}.
\end{equation}
Substituting the equilibrium controls, \eqref{wd*}, \eqref{eq: cosphi*_RA} and \eqref{eq: sinphi*_RA}, into the Hamiltonian, \eqref{eq: hamiltonian_RA}, and evaluating at $t_f$ gives
\begin{multline}\label{eq: H*_tf}
    \mathcal{H}_a^*(t_f) = 0 = 
    v_A\sqrt{1+\eta^2} -v_T(\eta \sin{\phi_T} +\lambda \cos{\phi_T})-1.
\end{multline}
% \begin{aligned}\label{eq: H*_tf}
%     \mathcal{H}_a^*(t_f) &=
%     v_A\sqrt{1+\eta^2} -v_T(\eta \sin{\phi_T} +\lambda \cos{\phi_T})-1.
%     \\
%     =0
% \end{aligned}
% \end{equation}
Solving \eqref{eq: H*_tf} gives
\begin{equation}\label{eq: eta}
    \begin{aligned}
\eta &=  \frac{ab \pm v_A\sqrt{a^2+b^2-v_A^2}}{v_A^2-a^2}, \\
    \end{aligned}
\end{equation}
where $a :=v_T\sin{\phi_T}$, and $b := (1+\lambda v_T\cos{\phi_T})$. 
% From assumption (A3), we know that the attacker will move in the positive $y$ direction, which implies $\sin{\phi_A^*}>0$. Based on this observation and \eqref{eq: sinphi*_RA}, we know $\eta>0$, 
If $\hat{y}_A < 0$, the attacker must move to the positive $y$ direction to reach the target, which implies $\sin{\phi_A^*}>0$. Based on this observation and \eqref{eq: sinphi*_RA}, we know $\eta>0$, and therefore the $+$ sign in \eqref{eq: eta} will be used. Likewise the $-$ sign will be used when $\hat{y}_A >0$.
% }.
% From assumption (A3) for initial state $\hat{y}_A<0$, $\sin{\phi_A}>0$, since $\phi_A \in [0, \pi]$. Therefore, $\eta>0$ due to \eqref{eq: sinphi*_RA}.  

% %=====================================%
\subsubsection{Solution Characteristics} 
The retrograde equilibrium kinematics \cite{Issacs1965} can be obtained by substituting the equilibrium controls, \eqref{wd*}, \eqref{eq: cosphi*_RA} and \eqref{eq: sinphi*_RA}, along with the adjoints into \eqref{eq: kinematics} which yields
% \begin{equation}\label{eq: xA_dot, yA_dot}
% \begin{aligned}
% \mathring{\hat{x}}^*_A &= 
%  \frac{\lambda v_A }{\sqrt{\eta^2 + 1}} -v_T\cos{\phi_T},
%  &
%  \mathring{\hat{y}}^*_A = 
% \frac{\eta v_A }{\sqrt{\eta^2 + 1}} -v_T\sin{\phi_T},
% \end{aligned}
% \end{equation}
\begin{equation}\label{eq: xA_dot, yA_dot}
\mathring{\hat{x}}^*_A = 
 \frac{\lambda v_A }{\sqrt{\eta^2 + 1}} -v_T\cos{\phi_T},\;\;\;
 \mathring{\hat{y}}^*_A = 
\frac{\eta v_A }{\sqrt{\eta^2 + 1}} -v_T\sin{\phi_T},
\end{equation}
with boundary condition, $\hat{y}_A(t_f)=0.$

Let $[X,Y]$ denote the relative position of the attacker with respect to the defender, i.e.,
\begin{equation}
    [X,Y] := [x_A-x_D, y_A-y_D] = [\hat{x}_A-\hat{x}_D, \hat{y}_A].
\end{equation}
% \edit{
Note that we have $\lambda = \sgn{(X)}$.
% }
Differentiating $X$ and $Y$ with respect to $t$, and manipulating the equations we have
% \begin{equation}\label{eq: m_Ra}
%     \begin{aligned}
%           m(\phi_A, w_D) := \frac{dY}{dX} =
%         %   \frac{dY}{dt}\cdot \frac{dt}{dX} = 
%           \frac{v_A\sin{\phi^*_A}-v_T\sin{\phi_T}}{v_A\cos{\phi^*_A}-v_T\cos{\phi_T} - w_D^*}
%     \end{aligned}.
% \end{equation}
\begin{equation}\label{eq: m_Ra}
    % \begin{aligned}
           m(\phi_A, \omega_D) := \frac{dY}{dX} =
        %   \frac{dY}{dt}\cdot \frac{dt}{dX} = 
           \frac{v_A\sin{\phi_A}-v_T\sin{\phi_T}}{v_A\cos{\phi_A}-v_T\cos{\phi_T} - \omega_D}.
    % \end{aligned}.
\end{equation}
Since both $\phi_A^*$ and $\omega_D^*$ are constant, 
% this implies 
the equilibrium trajectories of the system in the $XY$-plane are given by straight lines:
% \footnote{Note that \eqref{eq: m_Ra} is a generic expression of the slope for any given $\phi_A$ and $w_D$.
% }
\begin{equation} \label{eq: Value Trajectories Straight}
    Y = m^* X + C,
\end{equation}
where $m^* = m(\phi_A^*,\omega_D^*)$.
% with the slope given in \eqref{eq: m_Ra}.

\begin{figure}[t!]
    \centering
        \vspace{1mm}
    \includegraphics[width = \columnwidth ]{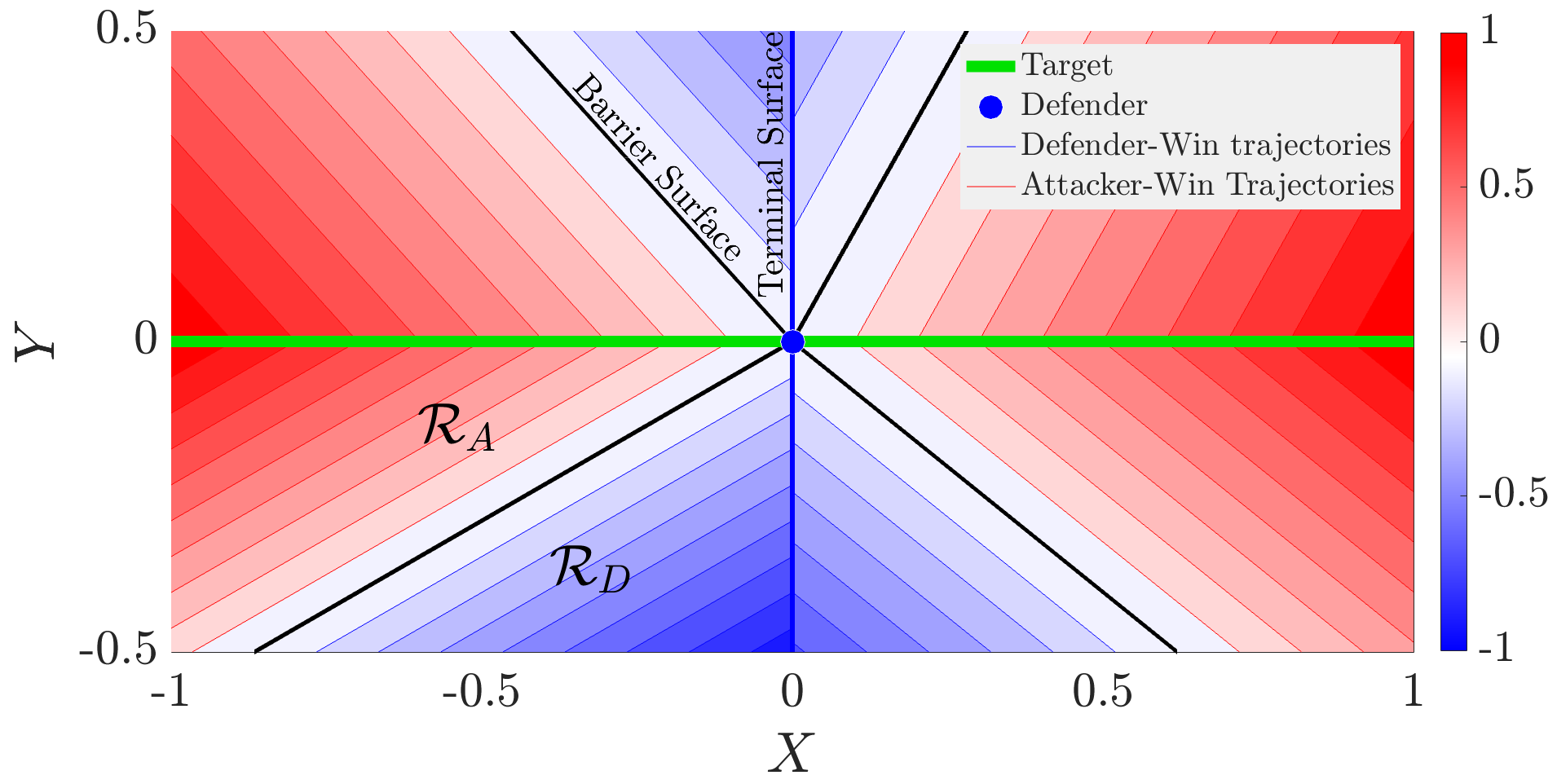}
    \caption{
    % \daigo{
    Equilibrium trajectories of the relative position vector $[X,Y]$ in infinite-length target scenario. The color indicates the Value of the game. The following parameters are used: 
    % }
    % for infinite length target, equilibrium trajectories of the players are straight lines. 
    % \sout{The barrier surface is indicated by the black lines that separates $\mathcal{R}_A$ and $\mathcal{R}_D$.} 
    % The red and blue lines indicate the level set of the Value of the game in $\mathcal{R}_A$ and $\mathcal{R}_D$, respectively, for 
    % \sout{a given initial configuration and} 
    % parameters: 
    $v_A = 0.7$, $v_T = 0.2$ and $\phi_T = 2\pi/3$.}  
    \label{fig:flow_field}
\end{figure}

The red solid lines in Fig.~\ref{fig:flow_field} present the equilibrium trajectories for $\hat{\mathbf{\x{}}} \in \mathcal{R}_A$. It can be seen that the terminal payoff in \eqref{payoff} is determined by the $X$ intercept of the state trajectory, which we denote by $X_f$.
The black solid lines indicate the critical case in which the attacker reaches the target at the time of capture with zero {miss-distance} (i.e., $X_f = 0$).
% \daigo{
Beyond this critical case, the region shown in blue is the defender-win region, which will be discussed in Sec.~\ref{sec: def-win}.
% }
% Beyond this critical case we obtain the set of trajectories shown in blue lines, indicates the defender-win region. 
% We shall derive the equilibrium strategies for this region in the subsequent sections.

% It can be seen that the terminal payoff in \eqref{payoff} is determined by the $X$ intercept of the state trajectory in the $XY$-plane, which we denote by $X_f$.
% The attacker tries to maximize $|X_f|$, while the defender tries to minimize it. 
% The result of these two conflicting intentions is the equilibrium trajectories in Figure~\ref{fig:flow_field}.

%%=============================================================
\begin{theorem}[Infinite-Length Target]\label{Theorem 1}
Consider the \textit{game of degree} with payoff given in \eqref{payoff}, and suppose the target is infinitely long.
% , i.e., spans the entire $\hat{x}$-axis.
Then the equilibrium state feedback control strategies are given by \eqref{wd*}, \eqref{eq: cosphi*_RA} and \eqref{eq: sinphi*_RA}.
% \begin{equation}
% \begin{aligned}\label{eq: eq_strategy_attacker}
% \left[\cos{\phi_A^*}, \sin{\phi_A^*}\right]
%  &= 
% \left[\frac{\sgn{\left(X\right)}}{\sqrt{\eta^2 + 1}} , \frac{\eta}{\sqrt{\eta^2 + 1}}\right],
% \end{aligned}
% \end{equation} 
% \text{where $\eta$ is given by \eqref{eq: eta}}, and
% \begin{equation}
%     \begin{aligned}
% w_D^* &= \sgn{\left(X\right)}.
% \end{aligned}
% \end{equation}
Moreover, the Value of the game is
\begin{equation}
    \begin{aligned}\label{Value_inf}
     V_a &= \sgn{\left(X\right)}\left( X - \frac{Y}{m^*}\right),
\end{aligned}
\end{equation}
where $m^*=m(\phi_A^*,\omega_D^*)$ with the expression given in \eqref{eq: m_Ra}.
% \footnote{This is a generic expression of the slope for any given $\phi_A$ and $w_D$.}
% \begin{equation}
%     m(\phi_A^*,w_D^*) := \frac{v_A\sin{\phi_A^*}-v_T\sin{\phi_T}}{v_A \cos{\phi_A^*-v_T\cos{\phi_T}-w_D^*}}.
%     \label{eq:optimal_slope}
% \end{equation}
\end{theorem}
\begin{proof} 
The players' strategies are derived using the first order necessary condition for optimality.
% Given the strategies, the relative position takes a straight line path in the $XY$-plane, $Y=mX+C$, with the slope given in \eqref{eq: m_Ra}.
% \eqref{eq:optimal_slope}.
% \footnote{This is a generic expression of the slope for any given $\phi_A$ and $w_D$. One can verify that $m$ in 
% \eqref{eq:optimal_slope} 
% matches $m_1$ in \eqref{eq: m_Ra}, when the agents use the equilibrium strategies.}
As discussed with Fig.~\ref{fig:flow_field}, the Value is given by the $X$ intercept of the equilibrium trajectory.
More specifically, the {miss-distance} is $X_f>0$ if the game starts in the positive $X$ region, whereas it is $-X_f>0$ if the game starts in the negative $X$ region. 
% (and thus the $X$ intercept is negative).
For a given initial condition $[X_0, Y_0]$, we have
\begin{equation}
\begin{aligned}
  C   &= Y_0 - m^* X_0.
\end{aligned}
\end{equation}
Substituting $C$ back into the equation and solving for the $X$ intercept gives:
\begin{equation}\label{eq:x_f}
    X_f = X_0 - \frac{Y_0}{m^*}.
\end{equation}
This completes the proof that \eqref{Value_inf} provides the Value of the game.
\end{proof}
%%======================================
% \subsection{Finite Length Target}
%%======================================
\subsection{Finite Length Target}
In the original problem, the endpoints of the target become important consideration.
Notice that there is always one endpoint that is relevant to the game: i.e., the one that the attacker may be able to reach without crossing $X=0$. We denote this endpoint as $\hat{\x{}}_E=[\hat{x}_E,0]^\top$, where
% Recalling the original problem where the length of the target is $L$, we denote the endpoint by $\hat{\x{}}_E = [\hat{x}_E,0]^\top$, in general, where $\hat{x}_E$ is defined as follows:
\begin{align}\label{eq: endpoint}
    \hat{x}_E &= (1+\sgn{(X)})\frac{L}{2}.
\end{align}
%  The defender, $D$, can travel only upto the end point of the target, and will follow the same strategy in equilibrium irrespective of the target length, since it only depends on the sign of $X$.
% The defender strategy will remain the same as it only depends on $X$. However, as the target is of finite length it can no longer travel beyond the end point of the target. 

The defender strategy will remain the same since it only depends on the relative position of the players, $X$. 
% However, due to the finite length of the target, it is no longer possible for it to travel beyond its endpoint.
However, the attacker's heading from Theorem~\ref{Theorem 1} is valid only if it intersects with the finite target.
% Likewise, the Attacker strategy is only valid if the attacker heading leads to \textit{breaching-point} on the finite target, otherwise the Attacker will miss the target.
% Alternatively, the attacker strategy is valid only if the attacker's heading leads to a \textit{breach-point} on the finite target, otherwise, the attacker will miss the target.
%
Let $\hat{\x{}}_B = [\hat{x}_B,0]^\top$ denote the point on the $\hat{x}$ axis
% $x$ axis of the target frame
that the attacker reaches following equilibrium strategies stated in Theorem~\ref{Theorem 1}:
% \eqref{eq: eq_strategy_attacker}:
\begin{equation}\label{hat(x)_B}
    \hat{x}_B := \hat{x}_A - \frac{\hat{y}_A}{m_B},\text{ where }m_B := \frac{v_A\sin{\phi^*_A}-v_T\sin{\phi_T}}{v_A\cos{\phi^*_A}-v_T\cos{\phi_T}}.   
\end{equation}

\begin{figure}[t!]
    \centering
        \vspace{1mm}
    \includegraphics[width = \columnwidth]{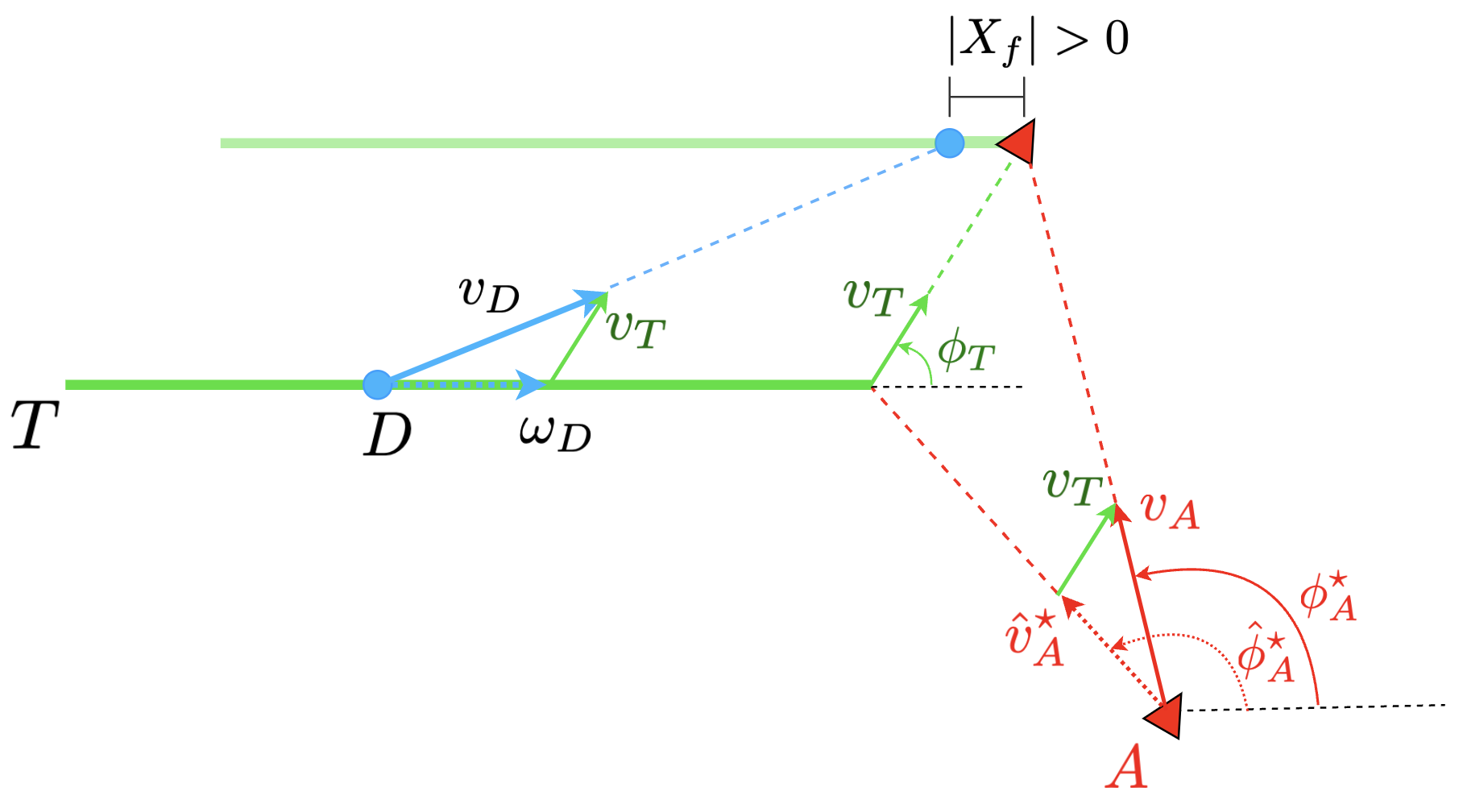}
    \caption{Attacker strategy in target frame and inertial frame. 
    When the strategy in Theorem~\ref{Theorem 1} fails to intercept the target, the attacker employs a heading to reach the endpoint of the target.}
    % Attacker deviates from the optimal heading, $\phi_A^*$, given by Theorem~\ref{Theorem 1} to reach the endpoint of the target.}
    \label{Fig: A_Traject}
\end{figure}

Now we can define two \textit{strategic regions} for the attacker-win game as follows:
\begin{itemize}
    \item $\mathcal{S}_{1a}$: $\hat{\x{}} \in \mathcal{R}_A$ and the strategy
    stated in Theorem~1
    % \eqref{eq: eq_strategy_attacker}
    is still valid for finite-length target case, given that the following condition holds:
\begin{equation}\label{eq:xB_condition}
    \hat{x}_B \in [0,L].
\end{equation}
    \item $\mathcal{S}_0$: $\hat{\x{}} \in \mathcal{R}_A$, however, \eqref{eq:xB_condition} does not hold.
\end{itemize}
% If \eqref{eq:xB_condition} does not hold, then 
In $\mathcal{S}_0$ the attacker must sacrifice the separation with the defender at $t_f$ and pick an aim point that actually intercepts the target.\footnote{Note that there is no incentive for the attacker to go around the endpoint and approach the target from the positive side, i.e., enter $\hat{y}_A>0$ region 
% \edit{because...}
because the attacker cannot improve the {miss-distance} as long as defender plays optimally.}
The aim point that achieves the least deviation from 
optimal heading, $\phi_A^*$,
% strategy stated in Theorem~1
% for infinite length target
% \eqref{eq: eq_strategy_attacker} 
is the endpoint $\hat{\x{}}_E$ as shown in Fig.~\ref{Fig: A_Traject}.

The attacker's heading angle in target frame for it to hit the endpoint $\hat{\x{}}_E$, is given by
% \begin{equation}\label{eq: cos_sin_fin}
% % \begin{aligned}
%     \left[\cos{\hat{\phi}_A^\finite}, \sin{\hat{\phi}_A^\finite} \right] &= \frac{1}{d_{EA}}
%     \left[\hat{x}_E-\hat{x}_A, -\hat{y}_A\right],
% % \end{aligned}    
% \end{equation}
\begin{equation}\label{eq: cos_sin_fin}
% \begin{aligned}
    \left[\cos{\hat{\phi}_A^\finite}, \sin{\hat{\phi}_A^\finite} \right] = \left[\frac{\hat{x}_E-\hat{x}_A}{\|\hat{\x{}}_E-\hat{\x{}}_A\|}, \frac{-\hat{y}_A}{\|\hat{\x{}}_E-\hat{\x{}}_A\|}\right].
% \end{aligned}    
\end{equation}
% where $d_{EA} := \sqrt{(\hat{x}_E-\hat{x}_A)^2+\hat{y}_A^2}$, is the distance from the attacker's position, $\hat{\x{}}_A$, to target's endpoint, $\hat{\x{}}_E$.
Note that we use the superscript $^\finite$ to denote the optimal strategies for the finite-length case.
Using the law of cosines 
% for the triangle $\triangle ABE$ in Fig.~\ref{Fig: A_Traject}, 
we obtain
\begin{equation}\label{eq: law of cosine} 
 v_A^2 = v_T^2+\hat{v}_A^2+2v_T\hat{v}_A\cos({ \hat{\phi}^\finite_A-\phi_T}).
\end{equation}
Solving for $\hat{v}_A$ yields
\begin{multline}\label{eq: v_Ahat_fin} 
    \hat{v}_A^\finite = 
    -v_T (\cos{\hat{\phi}_A^\finite}\cos{\phi_T}+\sin{\hat{\phi}_A^\finite}\sin{\phi_T})
    \\  \quad + \sqrt{v_A^2 - v_T^2 \left(\sin{\hat{\phi}_A^\finite}\cos{\phi_T}-\cos{\hat{\phi}_A^\finite}\sin{\phi_T}\right)^2}.
\end{multline}
Now we are ready to state the main theorem.

%%====================================================================================
\begin{theorem}[Finite-Length Target] \label{Theorem 2}
The equilibrium state feedback control strategy for the defender remains the same as stated in Theorem~\ref{Theorem 1}.
% is given by
% \begin{equation}
%     w_D^\finite = \sgn{\left(X\right)}.
% \end{equation}
The equilibrium state feedback strategy for the attacker is given in Theorem~\ref{Theorem 1}
% \eqref{eq: eq_strategy_attacker} 
if \eqref{eq:xB_condition} holds; 
otherwise, it is given by
\begin{equation}\label{eq:optimal_A_finite}
\begin{aligned}
    &\left[\cos{\phi_A^\finite}, \sin{\phi_A^\finite}\right] =\\
 & \frac{1}{v_A}\left[\hat{v}_A^\finite \cos{\hat{\phi}_A^\finite}+ v_T\cos{\phi_T},
 \hat{v}_A^\finite\sin{\hat{\phi}_A^\finite} 
  + v_T\sin{\phi_T}\right].
\end{aligned}
\end{equation}
where $\cos{\hat{\phi}_A^\finite}$, $\sin{\hat{\phi}_A^\finite}$, and  $\hat{v}_A^\finite$ are given by \eqref{eq: cos_sin_fin} and \eqref{eq: v_Ahat_fin} respectively. 
The Value of the game is
given by the expression in \eqref{Value_inf},
but with the slope $m^\star = m(\phi_A^\star,\omega_D^*)$ when \eqref{eq:xB_condition} does not hold.
% Note that based on condition \eqref{eq:xB_condition}, $\phi_A^*$ or $\phi_A^\star$ is used to compute $m$.
% and \eqref{eq:optimal_slope}. 
% Note that based on condition \eqref{eq:xB_condition}, either \eqref{eq: eq_strategy_attacker} or \eqref{eq:optimal_A_finite} is used in \eqref{eq:optimal_slope} to compute $m$.
\end{theorem}
% \begin{proof}
% Defender's strategy is same for both finite and infinite-length target, since it only depends on $X$.
 
% For the attacker, if $\hat{\x{}}\in \mathcal{S}_1$, the alternate heading ensures that it reaches the target by deviating the least amount possible from the equilibrium strategy given by \eqref{eq: eq_strategy_attacker}.

% From \eqref{eq:x_f}, we have seen that for any given initial conditions and $m$, 
% the expression in \eqref{Value_inf} gives the $X$ intercept (payoff).
% Furthermore, the expression in \eqref{eq:optimal_slope} provides the slope for any given strategy.
% Therefore, the generic expression \eqref{Value_inf} and \eqref{eq:optimal_slope} can be used for the finite-length strategy \eqref{eq:optimal_A_finite} as well.
% \end{proof}
The barrier surface is given by the zero-level set of the Value function \eqref{Value_inf}: 
% In the limiting case in which the {miss-distance} from the defender is zero, the surface 
\begin{align}
    \mathcal{S}_B = \{\hat{\x{}} \mid V_a = 0 \},
\end{align}
which separates the state space into attacker-win and defender-win regions, respectively,
\begin{equation}
    \begin{aligned}
        \mathcal{R}_A = \{\hat{\x{}} \mid V_a > 0 \}, \quad
        \mathcal{R}_D = \{\hat{\x{}} \mid V_a \leq 0 \}.
    \end{aligned}
\end{equation}
See Fig.~\ref{fig: equilibrium_flow_field} for the illustration of the barrier surface.
The closed form expression for the barrier surface will be discussed in Sec.~\ref{Section: V; GoD, GoK}.

%%====================================================================================
\section{DEFENDER-WIN SCENARIO}\label{sec: def-win}
In this section, we consider a {game of degree} for initial states in the defender-win region (i.e., $\hat{\x{}} \in \mathcal{R}_D$). We use the subscript $d$ to refer to the game in this region. 
% Defender wins the game if 
% % one of the following condition is being met:
% % \begin{enumerate}[I.]
%     (i) defender aligns with attacker: $\hat{x}_D(t_f) = \hat{x}_A(t_f)$, or
%     (ii) defender reaches the desired end of the target: $\hat{x}_D(t_f) = \hat{x}_E$, where $\hat{x}_E$ is defined in \eqref{eq: endpoint}.
% % \end{enumerate}
% % First condition is a part of terminal surface due to assumption (A2), and condition~{II} is also a part of terminal surface, since the attacker cannot reach the target with a non-zero \textit{miss-distance} without satisfying condition~{I}.
% The first condition is a part of the terminal surface due to assumption (A2). 
% The second condition is also a part of the terminal surface since the attacker cannot reach the target with a non-zero \textit{miss-distance} without satisfying first condition.
% The payoff is given by
We consider the following payoff function:
\begin{align}\label{eq: J_d}
    J_d(\hat{\x{}}_0, \omega_D, \phi_A) = \Phi_d(\hat{\x{}}_f) = -\sqrt{X_f^2+Y_f^2},
\end{align}
which is the negative of the distance between the attacker and the defender at terminal time. The negative sign is used to maintain the convention that the attacker (resp.~defender) is the minimizer (resp.~maximizer).
% The attacker tries to minimize and the defender seeks to maximize (\ref{eq: J_d}). 

%%====================================================================================
\subsection{Infinite Length target}
Similar to the attacker-win case, we start by looking into the infinite-length target case. Here the terminal condition is $\hat{x}_A = \hat{x}_D$. 
% The payoff function in \eqref{eq: J_d} reduces to the following:
If the target length is infinite, the payoff function in \eqref{eq: J_d} reduces to 
% attacker and defender play \textit{Game of Degree} over cost function given by
\begin{equation}\label{eq: Jd_inf}
    \phi_d(t_f) = -|Y_f|, \quad \text{where } Y_f = y_A(t_f) - y_D(t_f).
\end{equation}
The terminal constraint is given by
\begin{align}\label{eq: term_cond_RD_inf}
    \psi_d({\hat{\x{}}_f}) = \hat{x}_D(t_f) - \hat{x}_A(t_f) = 0.
\end{align}
Thus the terminal surface is defined by 
\begin{align}\label{eq: term_surf_RD_inf}
    \mathcal{S}_{T_{d, \text{inf}}} = \{\hat{\x{}} \mid \hat{x}_D(t_f) = \hat{x}_A(t_f)\}.
\end{align}
% Defender wins the game if 
%     (i) defender aligns with attacker: $\hat{x}_D(t_f) = \hat{x}_A(t_f)$, or
%     (ii) defender reaches the desired end of the target: $\hat{x}_D(t_f) = \hat{x}_E$, where $\hat{x}_E$ is defined in \eqref{eq: endpoint}.
% The first condition is a part of the terminal surface due to assumption (A2). 
% The second condition is also a part of the terminal surface since the attacker cannot reach the target with a non-zero \textit{miss-distance} without satisfying first condition.

The following theorem shows that the strategies remain the same as in the attacker-win scenario for infinite-length target.
\begin{theorem}[Infinite-Length Target]\label{Theorem 3}
The equilibrium state feedback control strategies for the attacker and the defender remain the same as as stated in Theorem~\ref{Theorem 1}, and the Value function is given by
\begin{align}\label{eq: V_d}
    V_d = \sgn{\left(Y\right)} \left( m^* X-Y \right),
\end{align}
where $m^* = m(\phi_A^*,\omega_D^*)$ is given in \eqref{eq: m_Ra}.
% \begin{equation}
%     m(\phi_A^*,w_D^*) := \frac{v_A\sin{\phi_A^*}-v_T\sin{\phi_T}}{{v_A \cos{\phi_A^*}-v_T\cos{\phi_T}-w_D^*}}.
%     \label{eq:optimal_slope_1}
% \end{equation}
\end{theorem}
\begin{proof}
This proof is based on the substitution of the proposed equilibrium strategies and Value function into the Hamiltonian-Jacobi-Isaacs (HJI) \cite{Issacs1965} equation:
% \begin{multline} \label{eq: HJI_d}
%         \min\limits_{w_D} \max\limits_{\phi_A} \left\{ l(\mathbf{\hat{\x{}}}, w_D, \phi_A) + \partial V_d /\partial t + V_\mathbf{\hat{\x{}}} \cdot f(\mathbf{\hat{\x{}}}, w_D, \phi_A)\right\} 
%     \\ = 0,
% \end{multline}
\begin{equation}\label{eq: HJI_d}
        \min\limits_{\omega_D} \max\limits_{\phi_A} \left\{ l(\cdot) + \partial V_d /\partial t + V_\mathbf{\hat{\x{}}} \cdot f(\cdot)\right\}  = 0,
\end{equation}
where the omitted function arguments are $(\cdot) = (\mathbf{\hat{\x{}}}, \omega_D, \phi_A)$, $V_{\hat{\x{}}}$ is the vector $[\partial V_d/\partial \hat{x}_D,  \partial V_d/\partial \hat{x}_A, \partial V_d/\partial \hat{y}_A]$, and $l$ represents an integral cost component. First, note that the cost, \eqref{eq: Jd_inf}, has no integral component, and thus $l = 0$. Also the proposed Value function, \eqref{eq: V_d}, is not an explicit function of time and thus $\partial V_d/\partial t = 0.$ The vector $V_{\hat{\x{}}}$ is obtained by differentiating \eqref{eq: V_d} with respect to each state:
\begin{align}
    V_{\hat{\x{}}}  = \sgn(Y)[-m^*,\; m^*,\; -1].
\end{align}
The (forward) equilibrium dynamics, $f$, are given by the negative of \eqref{eq: xA_dot, yA_dot}. Substituting all of these expressions into \eqref{eq: HJI_d} gives
\begin{multline*}
      \min_{\omega_D}\max_{\phi_A}\left\{\frac{\partial V_d}{\partial \hat{x}_D}\dot{\hat{x}}_D + \frac{\partial V_d}{\partial \hat{x}_A} \dot{\hat{x}}_A + \frac{\partial V_d}{\partial \hat{y}_A}\dot{\hat{y}}_A\right\} =
        \\ 
     -m^*(v_A\cos{\phi_A^*}-v_T\cos{\phi_T}-\lambda) +v_A\sin{\phi}_A^*-v_T\sin{\phi_T} \\
     =0.
\end{multline*}
Thus the proposed Value function is continuous and continuously differentiable, and it satisfies the HJI hyperbolic PDE.
\end{proof}

\subsection{Finite-Length target}
In this section we provide the equilibrium strategies for the finite length target for $\hat{\x{}} \in \mathcal{R}_D$. The defender strategy will remain the same for the finite length target. However, the defender is limited to move within the line segment. Therefore, the terminal surface is defined by 
% \begin{align}
%     \mathcal{S}_{T_d} = \{\hat{\x{}} \mid \hat{x}_D(t_f) = \hat{x}_A(t_f) \text{ or } \hat{x}_D(t_f) = \hat{x}_E \},
% \end{align}
\begin{align}
    \mathcal{S}_{T_d} = \{\hat{\x{}} \mid \hat{x}_D(t_{f_1}) = \hat{x}_A(t_{f_1}) \text{ or } \hat{x}_D(t_{f_2}) = \hat{x}_E \},
\end{align}
where, $t_{f_1}$ and $t_{f_2}$ are given by the time when attacker aligns with the defender, or the defender reaches the endpoint, respectively. The endpoint $\hat{x}_E$ is defined in \eqref{eq: endpoint}.
% Thus the terminal time 
% \begin{align}
%     t_f =\min\{t_{f_1},t_{f_2}\}.
% \end{align}
% For a given initial configuration $\hat{\x{}} \in \mathcal{R}_D$, based on the equilibrium strategy of the attacker, we can further divide it into following three strategic regions:
Note that $\hat{x}_D(t_{f_2}) = \hat{x}_E$ is part of the terminal surface since once the defender 
% approach towards the desired heading and 
reaches the desired endpoint, the attacker will no longer be able to reach the target without satisfying \eqref{eq: term_surf_RD_inf}. Thus the terminal time is
\begin{align}
    t_f =\min\{t_{f_1},t_{f_2}\}.
\end{align}

% \edit{
Let $\x{A}^*(t_{f_2}) = [x^*_A(t_{f_2}),y^*_A(t_{f_2})]^\top$ denote the point that the attacker will reach at $t_{f_2}$ following the strategy stated in Theorem~\ref{Theorem 3}:
\begin{align}
    x_A^*(t_{f_2}) = x_A + v_A\cos{\phi_A^*} \cdot t_{f_2}.
\end{align}
% }
% \sout{Let's consider that the attacker will reach $\x{A}^*(t_f) = (x^*_A(t_f),y^*_A(t_f))$, following the strategy stated in Theorem~\ref{Theorem 3}.} 
% \edit{
Also let us define a segment on $x$ axis bounded by $x_D$ and $x_E(t_{f_2})$ as follows:
\begin{equation}
    \mathcal{X}:=(x_D,x_E(t_{f_2})) \text{ or } (x_E(t_{f_2}),x_D).
\end{equation}
% } 
Now we can define three strategic regions for the defender-win game 
% \edit{
based on the location of $x_A$ and $x_A^*(t_{f_2})$
% } 
with respect to $\mathcal{X}$ as follows (also see Fig.~\ref{fig: equilibrium_flow_field}):
\begin{itemize}
    \item $\mathcal{S}_{1d}$: $\hat{\x{}} \in \mathcal{R}_D$, $x_A \in \mathcal{X}$, and the strategy stated in Theorem~\ref{Theorem 3}
    % \eqref{eq: eq_strategy_attacker}
    is still valid for finite-length target case, given that the following condition holds:
\begin{equation}\label{cond: xA*}
     x_A^*(t_{f_2}) \in \mathcal{X},
\end{equation}
% where $\mathcal{X}:=(x_D,x_E(t_{f_2})), \text{ or } (x_E(t_{f_2}),x_D,).$ 
% denotes the segment that the defender travels to reach the target endpoint.  
% is a set of $x$-coordinates of current and final position of the defender until it reaches the endpoint of the target in equilibrium.
% which means if the attacker's $x$-coordinate remains inside the defenders initial and final position the game will end by satisfying \ref{}
    \item $\mathcal{S}_2$: $\hat{\x{}} \in \mathcal{R}_D$ and $x_A \in \mathcal{X}
    % (x_D,x_E(t_{f_2}))
    $, however \eqref{cond: xA*} does not hold. 
    % Fig.~\ref{fig: S_3} illustrates attacker strategy in this region.
    \item $\mathcal{S}_3$:  $\hat{\x{}} \in \mathcal{R}_D$ and $x_A \notin \mathcal{X}
    % (x_D,x_E(t_{f_2}))
    $.
\end{itemize}
% For finite length target, attacker can use the same strategy if the  following condition holds:
% \begin{align}\label{cond: xA*}
%     x_A^*(t_f) \in (x_D,x_E(t_{f_2})),
% \end{align}
% where $t_{f_2}$ denotes the remaining time it takes for the defender to reach the endpoint\edit{:}
The time it takes for the defender to reach the endpoint is given by
% \edit{
% :
% }
% \begin{align}
%     t_{f_2} = |(\hat{x}_E-\hat{x}_D)/w_D^\star|=
%     |\hat{x}_E-\hat{x}_D|, \quad t_0 = 0.
% \end{align}
\begin{align}
    t_{f_2} = ||\hat{\x{}}_E-\hat{\x{}}_D||/|\omega_D^\star|=
    ||\hat{\x{}}_E-\hat{\x{}}_D||.
\end{align}
%  Note that the final time satisfies $t_f \leq t_{Df}$.
% Additionally, $x_E(t_{Df})$ is the $x$-coordinate of the
% \emph{target endpoint ($\x{E}$) at $t_{Df}$}. 
The coordinates of $\x{E}(t_{f_2})$ are given by the following:
% \begin{equation}\label{eq: xE(t_Df)}
%     \begin{aligned}
%         x_E(t_{Df}) &= x_E +  v_T\cos{\phi_T} \cdot t_{Df}, \\
%         y_E(t_{Df}) &= y_E +  v_T\sin{\phi_T} \cdot t_{Df}.
% \end{aligned}
% \end{equation}
\begin{equation}\label{eq: xE(t_Df)}
    \begin{aligned}
        x_E(t_{f_2}) &= x_E +  v_T\cos{\phi_T} \cdot t_{f_2}, \\
        y_E(t_{f_2}) &= y_E +  v_T\sin{\phi_T} \cdot t_{f_2}.
\end{aligned}
\end{equation}
% If \eqref{cond: xA*} holds attacker cannot improve it's payoff by unilaterally deviating from equilibrium strategy stated in Theorem~\ref{Theorem 3} and the game ends satisfying \eqref{eq: term_cond_RD_inf}. On the contrary, if \eqref{cond: xA*} does not holds, \eqref{eq: term_cond_RD_inf} will not satisfy, if attacker follows the same strategy. In this case attacker can improve from Theorem~\ref{Theorem 3} and seek for an alternate strategy. Based on the equilibrium strategy of the attacker, $\mathcal{R}_D$ can be further divided into the following strategic regions:

\begin{figure}[t!]
    \centering
        \vspace{2mm}
    \includegraphics[width = \columnwidth ]{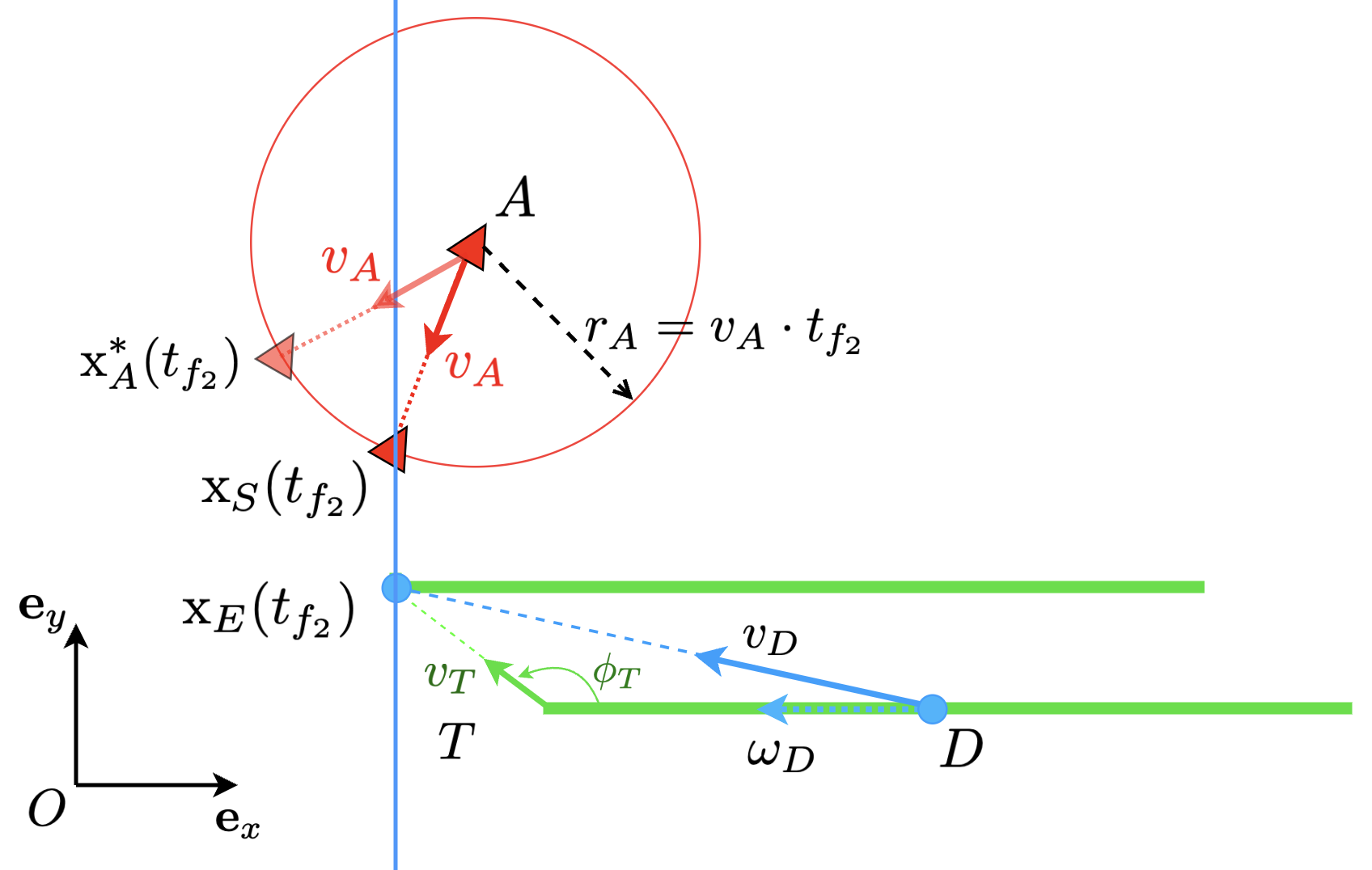}
    \caption{Attacker strategy for the initial state in $\mathcal{S}_2$. Attacker heading towards the {alignment point} $\x{s}(t_{f_2})$, where in equilibrium it aligns with the defender (i.e., $x_A = x_D$) at final time.  
    }
    \label{fig: S_3}
\end{figure}

If \eqref{cond: xA*} holds, then the attacker cannot improve its payoff by unilaterally deviating from the equilibrium strategy stated in Theorem~\ref{Theorem 3}, thus the game ends at $t_{f_1}$ by satisfying \eqref{eq: term_cond_RD_inf}. 
On the other hand, if \eqref{cond: xA*} does not hold, and if the attacker uses the strategy in Theorem~\ref{Theorem 3}, \eqref{eq: term_cond_RD_inf} will not be satisfied 
because the defender will reach the endpoint before it aligns with the attacker.
% In this case, attacker can improve on it's strategy from Theorem~\ref{Theorem 3}, and seek an alternate strategy. 
In this case, the attacker can choose an alternate heading to minimize the distance from the target endpoint at final time.\footnote{The suboptimality of the strategy in Theorem~\ref{Theorem 3} for $\hat{\x{}}_A \in \mathcal{S}_2$ is illustrated in Fig.~\ref{fig: S_3}.}
% \footnote{Note that Theorem~\ref{Theorem 3} provides optimal heading angle for the attacker when the target length is infinite, and the defender can travel indefinitely until it aligns with the attacker. For finite length target if $\hat{\x{}}_A \in \mathcal{S}_2$, the same strategy results in a suboptimal performance as illustrated in Fig.~\ref{fig: S_3}.}
% the attacker can improve the payoff by choosing to move towards $\x{S}(t_{f_2}).$}
% Based on the 
% equilibrium strategy adopted by 
% condition \eqref{cond: xA*} and initial position of the attacker, $\mathcal{R}_D$ can be further divided into the following strategic areas:
% \begin{align*}
%     \mathcal{S}_1:&= \{x_A \in (x_D, x_E(t_{Df})], x_A^*(t_f) \in (x_D,x_E(t_{Df}))\},
%         \\
%     \mathcal{S}_2:&= \{x_A \notin (x_D, x_E(t_{Df})],
%         \\
%     \mathcal{S}_3:&= \{x_A \in (x_D, x_E(t_{Df})], x_A^*(t_f) \notin (x_D,x_E(t_{Df})) \},
% \end{align*}
% \begin{align*}
%     \mathcal{S}_1:&= \{x_A \in (x_D, x_E(t_{f_2})], x_A^*(t_f) \in (x_D,x_E(t_{f_2}))\},
%         \\
%     \mathcal{S}_2:&= \{x_A \notin (x_D, x_E(t_{f_2})],
%         \\
%     \mathcal{S}_3:&= \{x_A \in (x_D, x_E(t_{f_2})], x_A^*(t_f) \notin (x_D,x_E(t_{f_2})) \}.
% \end{align*}
% If $\x{A}\in \mathcal{S}_3$, 
Specifically, the attacker will seek to align with the defender at time $t_{f_2}$ by deviating least amount from the optimal strategy given by Theorem~\ref{Theorem 3}.
We define this \emph{alignment point}, $\x{S}$, of the attacker and the defender at $t_{f_2}$ as follows:
\begin{equation}\label{eq: xS(t_Df)}
    \begin{aligned}
    x_S(t_{f_2}) &= x_E(t_{f_2}), 
    % = x_E +  v_T\cos{\phi_T} \cdot t_{Df}, 
    \\
    y_S(t_{f_2}) &= \begin{cases}
        \min\{y_1,y_2\}, \hspace*{\fill} &\text{if $Y>0$ and}, \\
        \max\{y_1,y_2\}, \hspace*{\fill} &\text{if $Y<0$.}
    \end{cases}      
    \end{aligned}
\end{equation}
Here $y_1$ and $y_2$ are given by
% the $y$-coordinates 
% of the intersection of the vertical \textit{terminal surface} and the circle drawn with a radius of maximum travel distance ($r_A$) of the attacker in $t_{Df}$ from a given initial position $\x{A}(x_A, y_A)$ shown in Figure~\ref{fig: S_3}:
\begin{equation}
    \begin{aligned}
    y_1 &= y_A + \sqrt{r_A^2-(x_A-x_A(t_{f_2}))^2}, \text{ and} \\
    y_2 &= y_A - \sqrt{r_A^2-(x_A-x_A(t_{f_2}))^2}.
    \end{aligned}
\end{equation}
where $r_A = v_A\cdot t_{f_2}$, is the distance traveled by the attacker by the time defender reach the endpoint $\x{E}(t_{f_2})$, and $[x_A, y_A]^\top$ is the attacker's initial position shown in Fig.~\ref{fig: S_3}.

\begin{theorem}[Finite-Length Target] \label{Theorem 4}
The equilibrium state feedback control strategy for the defender remains the same as stated in Theorem~\ref{Theorem 3}, and the equilibrium state feedback control strategy for the attacker is given in 
% remain the same as stated in
Theorem~\ref{Theorem 3} if  $\hat{\x{}} \in \mathcal{S}_{1d}$, and otherwise
% strategy in the defender-win region is given by
% \begin{multline}
%     \left[\cos{\phi_A^\star}, \sin{\phi_A^\star}\right] 
%     \\
%     = \begin{cases}
%     %  \left[\frac{\sgn{(X)}}{\sqrt{\eta^2 + 1}} , \frac{\eta}{\sqrt{\eta^2 + 1}}\right], \hspace*{\fill} &\text{if $\hat{\x{}} \in \mathcal{S}_1$,}
%     %  \\
%     \left[\frac{x_E(t_{D_f})-x_A}{\lVert \mathbf{\x{}}_E(t_{f_2})-\mathbf{\x{}}_A\rVert},
%     \frac{y_E(t_{D_f})-y_A}
%     {\lVert \mathbf{\x{}}_E(t_{f_2})-\mathbf{\x{}}_A\rVert}
%     \right], \hspace*{\fill} &\text{if $\hat{\x{}} \in \mathcal{S}_2$}, 
%     \\
%     \left[\frac{x_S(t_{D_f})-x_A}{\lVert \mathbf{\x{}}_S(t_{Df})-\mathbf{\x{}}_A\rVert},
%     \frac{y_S(t_{D_f})-y_A}{\lVert \mathbf{\x{}}_S(t_{Df})-\mathbf{\x{}}_A\rVert}
%     \right], \hspace*{\fill} &\text{if $\hat{\x{}} \in \mathcal{S}_3$}.
%     \end{cases}
% \end{multline}
\begin{multline}
    \left[\cos{\phi_A^\star}, \sin{\phi_A^\star}\right] 
    \\
    = 
    \begin{cases}
    \left[\frac{x_S(t_{f_2})-x_A}{\lVert \mathbf{\x{}}_S(t_{f_2})-\mathbf{\x{}}_A\rVert},
    \frac{y_S(t_{f_2})-y_A}{\lVert \mathbf{\x{}}_S(t_{f_2})-\mathbf{\x{}}_A\rVert}
    \right], \hspace*{\fill} &\text{if $\hat{\x{}} \in \mathcal{S}_2$},
    \\[4pt]
    \left[\frac{x_E(t_{f_2})-x_A}{\lVert \mathbf{\x{}}_E(t_{f_2})-\mathbf{\x{}}_A\rVert},
    \frac{y_E(t_{f_2})-y_A}
    {\lVert \mathbf{\x{}}_E(t_{f_2})-\mathbf{\x{}}_A\rVert}
    \right], \hspace*{\fill} &\text{if $\hat{\x{}} \in \mathcal{S}_3$}.
    \end{cases}
\end{multline}
where $\x{E}(t_{f_2})$ and $\x{S}(t_{f_2})$ are given by \eqref{eq: xE(t_Df)} and \eqref{eq: xS(t_Df)}, respectively.
If $\hat{\x{}} \in \mathcal{S}_2 \cup \mathcal{S}_3$, the Value function is given by 
\begin{multline} \label{eq: V_d_S2-S3}
     V_d = -\bigl\{\left(X+(v_A\cos{\phi_A^\star}-v_T\cos{\phi_T}-\omega_D^\star\right)\cdot t_{f_2})^2 
     \\
     + \left(Y+(v_A\sin{\phi_A^\star}-v_T\sin{\phi_T}\right)\cdot t_{f_2})^2\bigr\}^{1/2},
\end{multline}
otherwise, it is given in \eqref{eq: V_d}.
\end{theorem}

\begin{proof}
If $\hat{\x{}} \in \mathcal{R}_D$, the attacker cannot reach the target using the strategy in equilibrium. Therefore, the attacker seeks to minimize the distance at terminal time. If $\hat{\x{}} \in \mathcal{S}_2$
(resp. $\hat{\x{}} \in \mathcal{S}_3$)
% $\mathcal{S}_3$, 
the closest point from the target at final time is given by $ \x{E}(t_{f_2})$ (resp. $\x{S}(t_{f_2})$). 
% and \x{S}(t_{f_2})$ respectively.  

The relative position of the attacker to the defender in $x$ and $y$ direction at $t_{f_2}$ is given by
\begin{equation}
    \begin{aligned}
     X_f &=    
    X-(v_A\cos{\phi_A^\star}-v_T\cos{\phi_T}-\omega_D^\star)\cdot t_{f_2}, \\
     Y_f &=   
    Y + (v_A\sin{\phi_A^\star}-v_T\sin{\phi_T})\cdot t_{f_2}. 
    \end{aligned}
\end{equation}
Thus at final time, the distance between the players is given by \eqref{eq: V_d_S2-S3}.
% \begin{multline}
%     % \sqrt{X_f^2+Y_f^2}
%     % \\
%   J_d = -\bigl\{(X+(v_A\cos{\phi_A^\star}-v_T\cos{\phi_T}-w_D)\cdot t_{f_2})^2  \\ 
%     + (Y + \left(v_A\sin{\phi_A^\star}-v_T\sin{\phi_T})\cdot t_{f_2}\right)^2\bigr\}^{1/2}.
% \end{multline}
\end{proof}
%%====================================================================================
\section{\uppercase{Game of Kind}}\label{Section: V; GoD, GoK}
\begin{figure}[t!]
    \centering
        \vspace{1mm}
    \includegraphics[width = \columnwidth ]{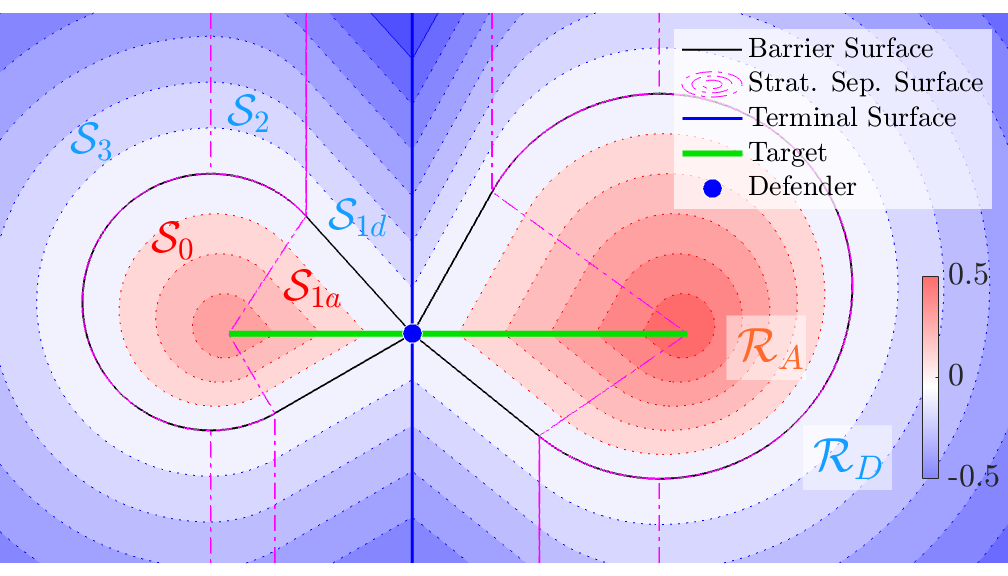}
    \caption{Illustration of level set of the Value of the game along with different strategic regions for equilibrium attacker strategy for $\hat{x{}}_D = 0.4$ and $v_A = 0.7$, $v_T = 0.2$ and $\phi_T = 2\pi/3$.}
    \label{fig: equilibrium_flow_field}
\end{figure}

Following our previous analyses
% of attacker-win and defender-win scenarios 
in Sec.~\ref{sec: att-win} and Sec.~\ref{sec: def-win}, Theorem~\ref{Theorem 1}-\ref{Theorem 4} provides the equilibrium strategies and Value function for the {game of degree}. Figure~{\ref{fig: equilibrium_flow_field}} shows the attacker and defender-win regions along with different strategic regions based on the equilibrium attacker strategy. The terminal surface from the defender's position segregates the state space into two regions, and the defender's strategy depends on which side the attacker resides in. 
% The barrier surface separates the state space based on the winning regions of the players.

 In Fig.~{\ref{fig: equilibrium_flow_field}}, the barrier surface for the {game of kind} is indicated by the black line which divides the state space into defender-win and attacker-win regions.
% \subsection{Closed form expression of the barrier surface}
% This section provides an analytical expression for the barrier surface. 
It is composed of two sections: a linear section and a circular section. 
The linear section is given by 
% $\mathcal{B}_l :=\{\hat{\x{}}\mid Y = mX\}$.
\begin{equation*}
    \begin{aligned}
      \mathcal{S}_{B,\text{linear}} :=\{\hat{\x{}}\;|\;Y &= m^*X\}.
    \end{aligned}
\end{equation*}
The circular section is denoted by $\mathcal{S}_{B,\text{circular}}$, whose center is at $\x{C} =[x_c,y_c]^\top$: 
\begin{equation*}
    \begin{aligned}
     x_c &=  x_E+v_T\cos{\phi_T}\cdot t_{f_2},\\
     y_c &= y_E+v_T\sin{\phi_T}\cdot t_{f_2},
    \end{aligned}
\end{equation*}
% = [x_E+v_T\cos{\phi_T}\cdot t_{Df},y_E+v_T\sin{\phi_T}\cdot t_{Df}]^\top$, is the center, 
% and 
and the radius is
$r_c:= v_A\cdot t_{f_2}$. 
% If \eqref{eq:xB_condition} holds for 
% \eqref{eq: eq_strategy_attacker}
% strategy given in Theorem~\ref{Theorem 1}, the barrier is given by 
% $\mathcal{B}_l$, 
% otherwise $\mathcal{B}_c$. Combining these two sections results in a complete barrier surface  for the {game of kind}.
% that separates the attacker-win and defender-win region.
The transition between the circular and linear part occurs at critical points where $\hat{x}_B=0$ or $\hat{x}_B=L$ (recall condition \eqref{eq:xB_condition}).
% \subsection{Equilibrium Strategies and Value of the Game}
% From the previous analysis of the attacker-win and defender-win scenarios in Section~III and Section~IV, Theorem~1-4 provides the equilibrium strategy for the entire $xy$-plane. Figure~{\ref{fig: equilibrium_flow_field}} shows the attacker and defender-winning regions along with different strategic regions based on the equilibrium attacker strategy. Note that, the terminal surface from the defenders position separates the state space into two regions, and defenders strategy depends on which side the attacker resides. 

% Following our previous analyses of attacker-win scenarios in Section~III and Section~IV, Theorem~1-4 provides the equilibrium strategy for the entire $xy$-plane. Figure~{\ref{fig: equilibrium_flow_field}} shows the attacker and defender-winning regions along with different strategic regions based on the equilibrium attacker strategy. The terminal surface from the defender's position segregates the state space into two regions, and the defender's strategy depends on which side the attacker resides on. 
% % The barrier surface separates the state space based on the winning regions of the players.
% Furthermore, the barrier surface divides the state space into defender-win and attacker-win regions.
%%======================================
\section{SIMULATIONS}
In this section, attacker-win and defender-win scenarios are illustrated for the following parameters:
$v_A=0.7$, $v_T=0.2$, $\phi_T=2\pi/3$  and $L=1$.\footnote{
The animated version of the simulations can be found online at \texttt{https://youtu.be/WJUvbmYj3AU}.}
For all the following examples,
% defender initial state remains the same, which is given by 
$\x{D}(t_0) = [0.4, 0.0]$, and the target frame coincides with the inertial frame at $t_0$. 

\begin{figure}[tpb!]
    \centering
        \vspace{1.5mm}
    \includegraphics[width =
    % \columnwidth
    7.5cm
    ]{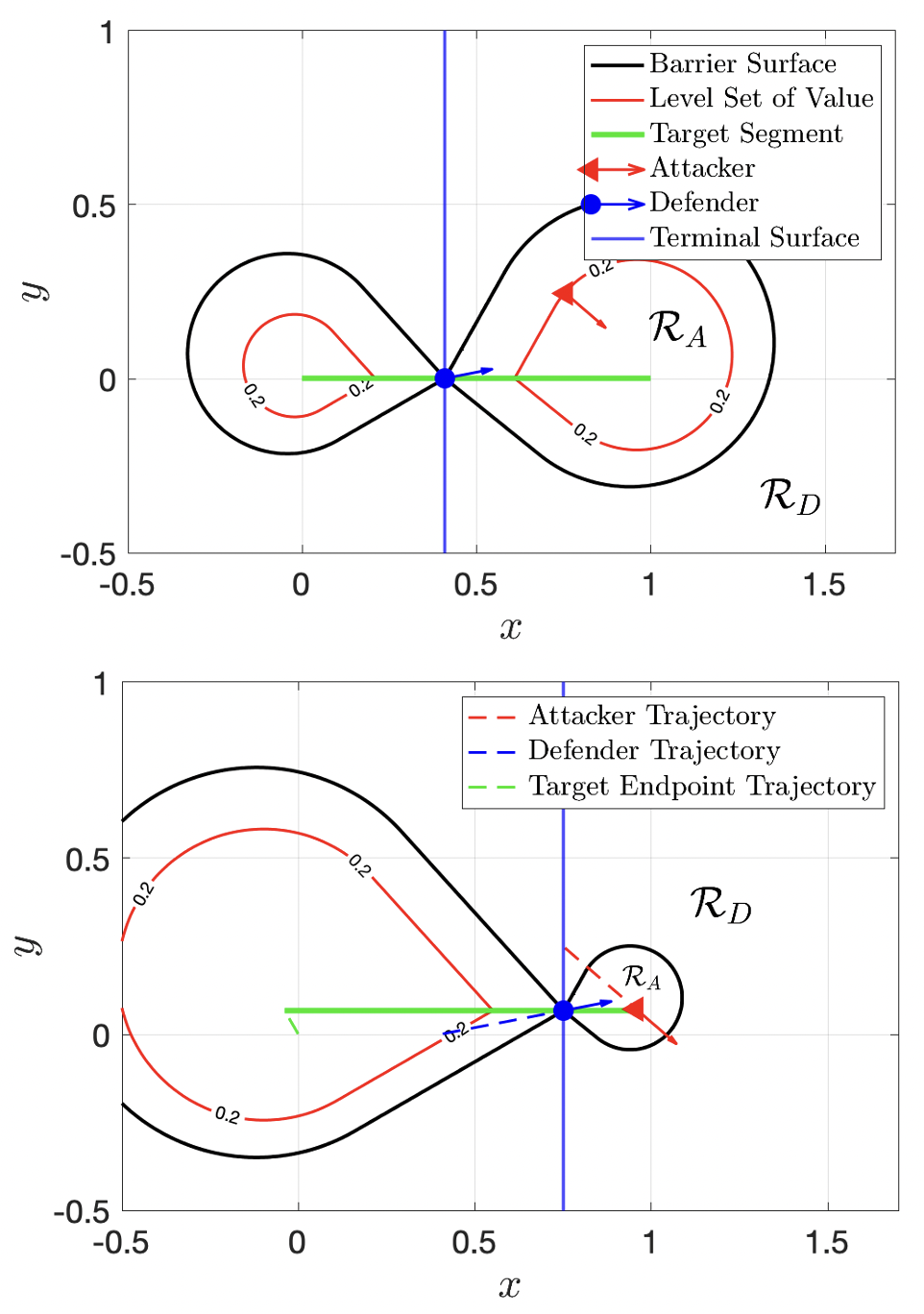}
    \caption{Example of attacker-win case, game state at time $t_0$ and $t_f$ with equilibrium strategies.}
    \label{fig: sim_RA}
\end{figure}

% \subsection{Attacker-Win Scenario}
In Fig.~\ref{fig: sim_RA}, $\x{A}(t_0) = [0.75,0.25]^\top \in \mathcal{S}_0$, and $\x{0} \in \mathcal{R}_A$.
% , which is inside the attacker-win region. 
% For the given equilibrium strategies attacker wins the game by reaching the target with a positive \textit{miss-distance} from the defender. 
% Both player employs their respective equilibrium strategies. 
This initial condition gives $V_a = 0.2$.
Under the equilibrium strategies on both players, the attacker stays on this level set throughout the game and reaches the endpoint of the target with $J_a=0.2$.
% At final time, $t_f$, attacker reaches the endpoint of the target and
% the Value of the game, $V_a = 0.2$, shown by the level set in the figure, which is constant throughout the game.
% \subsection{Defender-Win Scenario}
% Figure~7 shows the case where defender wins the game where, $\x{A}(t_0) = [0.05,0.50]^\top \in \mathcal{R}_D$. 
% % which is inside the defender-win region. 
% Both player employs their respective equilibrium strategies and the game ends when defender aligns with the attacker at terminal time. The Value of the game, $V_d = -0.167$, shown by the level set in the figure, which is constant throughout the game.
%

\begin{figure}[tpb!]
    \centering
        \vspace{1.5mm}
    \includegraphics[width =
    % \columnwidth
    7.5cm
    ]{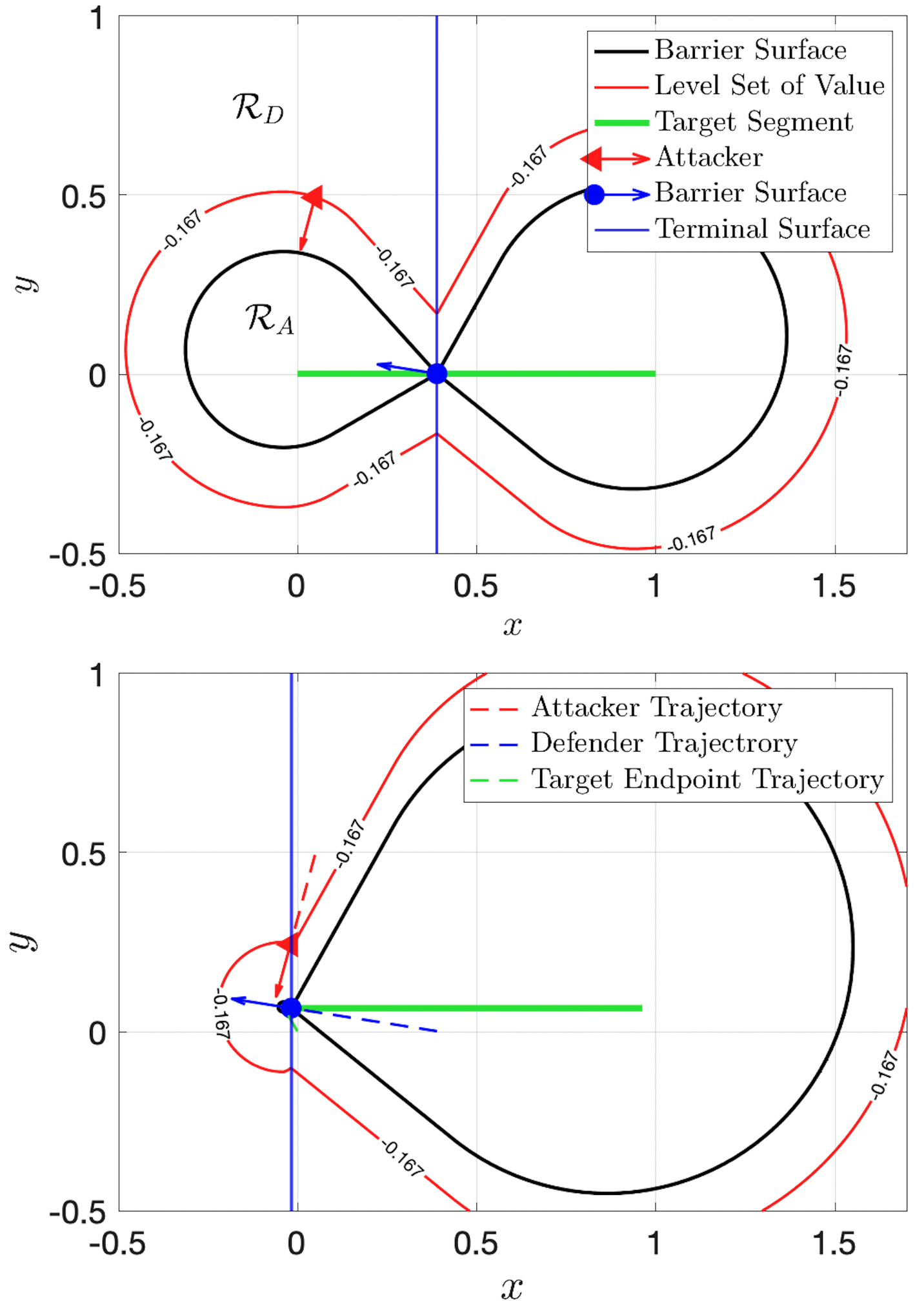}
    \caption{Example of defender-win case at initial and final state of the game with equilibrium strategies.}
    \label{fig: sim_RD}
\end{figure}

In Fig.~\ref{fig: sim_RD}, $\x{A}(t_0) = [0.05,0.50]^\top \in \mathcal{S}_2$, and $\x{0} \in \mathcal{R}_D$.
At final time, $t_f$, attacker reaches the {alignment point} and defender reaches the endpoint of the target.
The Value of the game is, $V_d = -0.167$. The negative Value indicates the defender win case as oppose to positive Value for the attacker win game.
% , shown by the level set in the figure, which is constant throughout the game.
% \subsection{Rational Defender vs Irrational Attacker}
% Figure~\ref{fig: sim_suboptimal} shows the case where defender wins the game against an irrational attacker. The attacker plays with a naive strategy, i.e., it moves strait towards the target following the minimum distance path instead of equilibrium strategy, in contrast the defender plays with the equilibrium strategy. Initial state $\x{A} = [0.85, 0.48]^\top$, which is inside the attacker-win region at $t_0$, however, system state shifted inside the defender-win region at some time $t_f \leq t > t_0$, due to the \textit{sub-optimal} strategy of the attacker. The potential Value of the game for equilibrium strategy (+0.036) and the actual payoff (-0.037) is also shown in the figure.

\begin{figure}[tb!]
    \centering
        \vspace{1mm}
    \includegraphics[width = 
    \columnwidth
    % 7.5cm
    ]{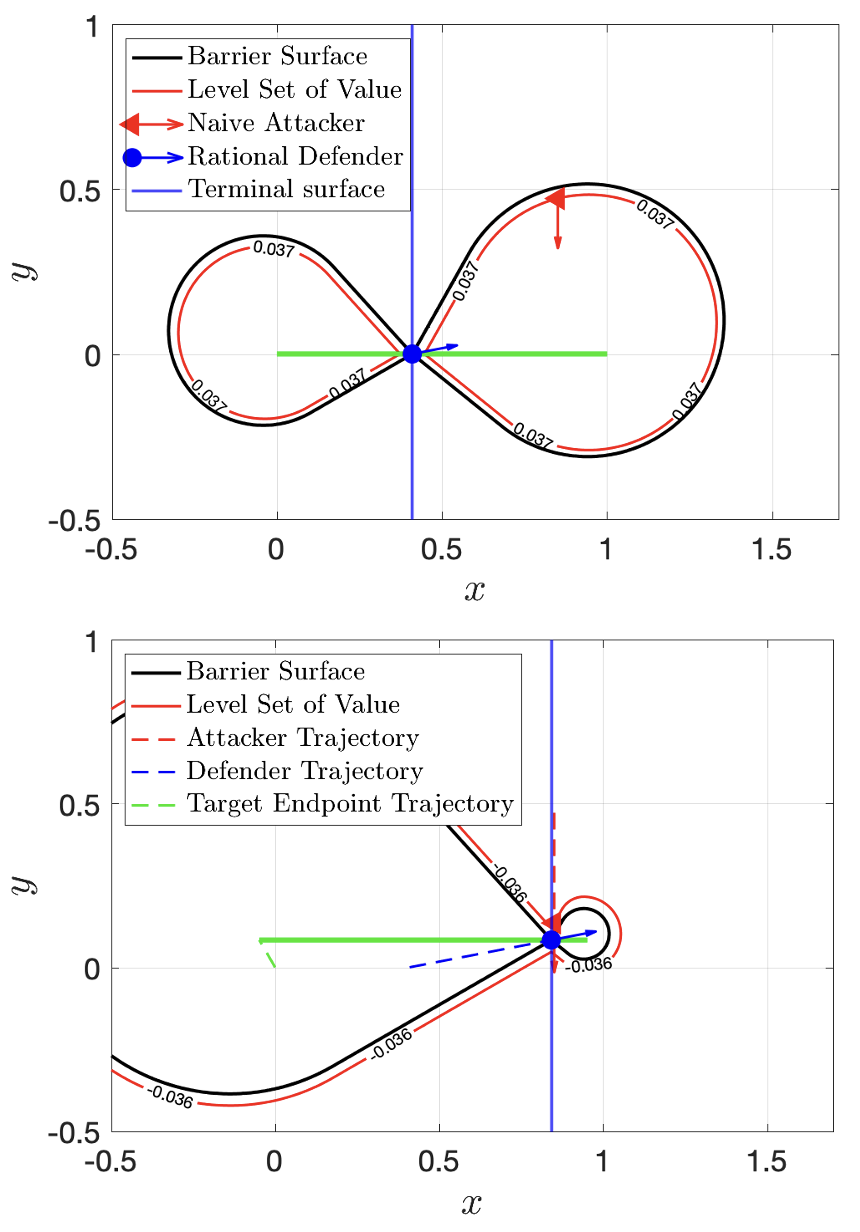}
    \caption{Example of defender-win case where defender wins against a naive attacker that employs a sub-optimal strategy.} 
    % Initial state $\hat{\x{}}_0 \in \mathcal{R}_A$, and final state  $\hat{\x{}}_f \in \mathcal{R}_D$}.
    % }
    \label{fig: sim_suboptimal}
\end{figure}

Figure~\ref{fig: sim_suboptimal} illustrates a scenario in which the attacker looses the game starting from a winning position when it employs a sub-optimal strategy.
At time $t_0, \x{A} = [0.85, 0.48]^\top \in \mathcal{S}_0$, and $\x{0} \in \mathcal{R}_A$. 
% resides inside the attacker-win region at $t_0$.
In equilibrium, the attacker will seek the endpoint of the target and wins the game. However, in this example attacker employs a sub-optimal strategy (i.e., move straight towards the target).
As a result, the state shifted inside the defender-win region at some time $t_f \geq t > t_0$ and the game ends with capture. 
% eventually captured by the defender at $t_f$. 
The figure also depicts the potential Value of the game under equilibrium strategy $(+0.036)$ at $t_0$, and the actual payoff $(-0.037)$ at $t_f$.
\section{CONCLUSIONS}
% In this paper, we have extended the target guarding problem for any given direction of translation of the target.
% We have derived the equilibrium strategies of the players and Value of the game for infinite length target and used the result to solve the problem for a finite length target.
% To solve for \textit{Game of Kind}, we have provided the expression of the barrier surface both in numerical and close form to separate the attacker and defender win region. Additionally, we have redefined the payoff for the defender win region in terms of distance between the players.
% Furthermore, we have provided examples of defender and attacker win scenario using optimal strategies and how unilaterally deviating from these strategies will result in a different outcome for the players. 
% Finally, future works may include more practical shape of the targets involving multiple defenders and attackers. The information structure and the dynamics of the players can be changed to fit into the real world scenarios.
%
In this paper, we address the problem of defending a non-maneuverable translating target. By determining players' equilibrium strategies and the Value of the game for an infinite-length target, we were able to leverage those results to the original problem with finite-length target. As a solution to the {game of kind}, we provide expressions of the barrier surface both in the numerical form and in an analytical form. 
% for separating the attacker-win region from the defender-win region. 
% Furthermore, we have redefined the payoff for the defender-win region based on the terminal distance between the players. 
In addition, we provided examples of defender-win and attacker-win scenarios using optimal strategies, and we examined how unilateral deviation from these strategies would affect the outcome of the game. Future works may include more practical shapes of the targets involving multiple defenders and attackers. Information structure and dynamics of the players can be adapted to fit real-world situations.
% \typeout{} 
\bibliographystyle{IEEEtran}
\bibliography{Citation}
% \bibliography{cite_concise}
\end{document}